\documentclass[fleqn,usenatbib]{mnras}

\usepackage{newtxtext,newtxmath}

\usepackage[T1]{fontenc}

\DeclareRobustCommand{\VAN}[3]{#2}
\let\VANthebibliography\thebibliography
\def\thebibliography{\DeclareRobustCommand{\VAN}[3]{##3}\VANthebibliography}

\usepackage{graphicx}	
\usepackage{amsmath}	
\usepackage[nopatch]{microtype}
\usepackage{booktabs}
\usepackage{blindtext} 
\usepackage{comment}
\usepackage{makecell}
\usepackage{multicol}
\usepackage{float}
\usepackage{lscape}
\usepackage{braket}
\usepackage{color}
\usepackage{algorithm}
\usepackage{siunitx}
\usepackage{float}
\usepackage{aas-macros}
\usepackage{bigints}
\usepackage{braket}
\usepackage{mathtools,esdiff}
\usepackage{booktabs}
\usepackage{mwe}
\usepackage{subcaption}
\usepackage{caption}
\usepackage{makecell}
\usepackage{multicol}
\usepackage{float}
\usepackage{lscape}
\usepackage[export]{adjustbox}
\usepackage{blindtext}
\usepackage[font=small,labelfont=bf]{caption}
\setcellgapes{10pt}

\DeclareMathOperator{\arccosh}{arccosh}
\DeclareMathOperator{\arctanh}{arctanh}

\title[Gravitational lensing via the Fox $H$-function]{Gravitational lensing by a generalised-NFW halo via the Fox $H$-function and its application to the super-NFW}

\author[D. A. Torres-Ballesteros and L. Casta\~neda]{
Daniel Alexdy Torres-Ballesteros\thanks{E-mail:daatorresba@unal.edu.co} and Leonardo Casta\~neda
\\
Observatorio Astron\'omico Nacional, Universidad Nacional de Colombia, Carrera 30 Calle 45-03, P.A. 111321 Bogot\'a, Colombia
}

\date{Accepted XXX. Received YYY; in original form ZZZ}

\pubyear{\the\year{}}

\begin{document}
\label{firstpage}
\pagerange{\pageref{firstpage}--\pageref{lastpage}}
\maketitle

\begin{abstract}
We present an analytical framework for a family of axisymmetric gravitational lenses, in which we express the lensing properties in terms of the Fox $H$-function. We apply this framework to a generalised-NFW (gNFW) profile, where we provide the power series representation of the Fox $H$-functions involved, and explore their performance and accuracy. From these power series we show that the corresponding Fox $H$-functions reduce to simple expressions in terms of the Gauss hypergeometric function. We apply these results to the particular case of the super-NFW (sNFW) profile, obtaining simpler expressions, this time in terms of complete elliptic functions (which are easier to work with). When the number of images formed is maximum, the sum of their signed magnifications denoted as $I$, is constant for several lenses. We study its behaviour for the sNFW, NFW and Hernquist lenses, and show that for a fixed $\kappa_0$ (characteristic convergence), in general, $I$ is not constant ($I_{\text{min}}\leq I \leq I_{\text{max}}$), as it exhibits a strong dependence on the source position inside the radial caustic. The boundaries depend on $\kappa_0$ (and so does the average $\braket{I}$). Our numerical experiments suggest that for these lenses $I_{\text{min}}\to 1$ as $\kappa_0$ increases, and $I\to 1$ as $\kappa_0\to \infty$. Additionally,  $I$ is constant only for a specific $\kappa_0$, which is different for each model.
\end{abstract}

\begin{keywords}
gravitational lensing: strong -- gravitational lensing: weak -- galaxies: haloes -- methods: analytical.
\end{keywords}



\section{Introduction}\label{sec: intro}
In our current understanding  of the Universe, the dark sector contributes to about $95\%$ of its content. The two dark components, known as dark energy and dark matter, are fundamental to our $\Lambda$CDM paradigm. One success of this paradigm is its ability to account for primordial density fluctuations that are amplified by the expansion of the Universe and, when they reach a critical value, collapse into dark matter haloes. Understanding how these dark matter haloes form and evolve is one of the most important tasks in cosmology and astrophysics. Gravitational lensing stands as a fundamental tool to unveil the nature of this dark sector at different scales (see  e.g., \cite{bartelmann2001wl_review,munshi2008weak_lensing_survey,hoesktra2008weak_lensing, ellis2010lensing_review, huterer2010weak_lensing_review, massey2010darkmatter_lensing, umetsu2020wl_cluster, natarajan2024sl_clusters}). 

When we attempt to model astrophysical systems such as galaxies, galaxy groups, or galaxy clusters as gravitational lenses, fitting observational data is required. This can be done, for instance, using MCMC algorithms (or algorithms of similar nature), which require repeated evaluation of the lensing properties for the different components of such systems (e.g., \cite{jullo2007lenstool, oguri2010PASJgrafic}). This process is computationally expensive and time-consuming, especially when numerical integration is necessary (which scales with the complexity of the mass distributions). If the analytical properties of a required gravitational lens model are unavailable, an efficient alternative to numerical integration is to develop fast calculations methods. For example, \cite{oguri2021fast} proposed the superposition of cored steep ellipsoids to compute the lensing properties for elliptic NFW and elliptic Hernquist lenses, an approach that he implemented in \textsc{glafic} \citep{oguri2010PASJgrafic}. Therefore, we aim to derive analytical expressions for the lensing properties of the generalised-NFW (gNFW)   profile proposed by \cite{evansGNFW2006}, a spherically symmetric profile defined as
\begin{equation}\label{eq: gnfw density}
\rho(r)=\dfrac{(n-2)\phi_0}{4\pi Gr_0^2}\dfrac{1}{(r/r_0)(1+r/r_0)^{n-1}},
\end{equation}
where $r_0$ is a length scale, $\phi_0$ is the depth of the potential well, and $n>2$. For this mass distribution the logarithmic slope takes the form
\begin{equation}
\dfrac{d\ln{\rho}}{d \ln{r}} = -1+\dfrac{(1-n)(r/r_0)}{\left(1+r/r_0\right)},
\end{equation}
where we can show that the gNFW satisfies $\rho(r)\propto r^{-1}$ as we approach to the centre ($r\ll r_0$), while it falls off like $\rho(r)\propto r^{-n}$ when $r\gg r_0$.

\begin{figure}
\centering
\includegraphics[width=0.9\linewidth]{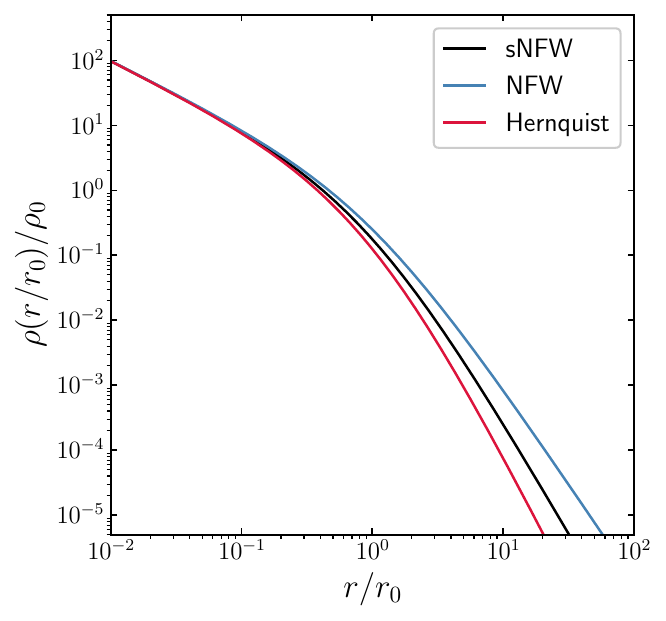}
\caption{Normalised volumetric mass density given for the sNFW (black), NFW (blue), and Hernquist (red) profiles. Here, $\rho_0=(n-2)\phi_0/(4\pi Gr_0^2)$ is the characteristic density.}
\label{fig: volumetric mass density}
\end{figure}

This family of mass distributions includes as special cases the NFW for $n=3$ \citep{nfw1997}, Hernquist for $n=4$ \citep{Hernquist1990model}, and super-NFW (sNFW) for $n=7/2$ \citep{lilley2018snfw}. In Fig. \ref{fig: volumetric mass density} we can see that for these three profiles the density changes at the same rate towards the centre (as any profile described by the gNFW). Towards the outskirts, however, the sNFW falls off faster than the NFW but slower than the Hernquist profile (precisely one of the motivations for defining the sNFW). \cite{lilley2018snfw} showed that the sNFW is at least as good as the NFW in reproducing haloes extracted from $N$-body simulations, with the added advantage that the sNFW has a finite total mass. Additionally, the sNFW provides a better approximation to the surface brightness of elliptical galaxies and bulges in disc galaxies than the Hernquist profile. It can also be easily extended to distorted profiles beyond spherical symmetry (e.g., flattened or triaxial profiles), as it facilitates the constructions of a basis for a biorthogonal expansion of the potential-density pair. For further details, we refer the reader to \cite{lilley2018snfw, lilley2018expansion} and the references therein.

For the description of the gNFW \eqref{eq: gnfw density} as a gravitational lens, the direct evaluation of the required integrals (see Sec. \ref{sec: gravitational lensing basics}) is not necessarily evident or trivial. Fortunately, the Mellin transform (see Appendix \ref{appendix: mellin}) and the Fox $H$-function (see Appendix \ref{appendix: fox}), working together, provide an elegant and powerful combination that allows us to achieve our goal. Within the astronomical context, this combination has been successfully applied, for instance, to the deprojection of the Sérsic surface brightness model \citep{baes2011sersic1,baes2011sersic2}, and to express the gravitational lensing properties of the Einasto profile \citep{retana-montenegro2012einasto-analytical, retana-montenegro2012einasto-weak} and the Dekel-Zhao profile \citep{dekel-zhao2020}.

This work is structured as follows: In Sec. \ref{sec: gravitational lensing basics} we establish notation and present the necessary elements of gravitational lensing for axisymmetric lenses. Then, in Sec. \ref{sec: general result}, we derive the lensing properties for a particular family of lenses via the Fox $H$-function, which we apply to the gNFW profile in Sec. \ref{sec: gNFW lensing}. In Sec. \ref{sec: sNFW lensing}, we focus on the sNFW, where we derive its analytical lensing properties. Also, for the sNFW, NFW and Hernquist lenses, when the number of images produced by these lenses reaches its maximum (three in this case), we study the behaviour of the sum of their signed magnifications. Finally, a summary and conclusions are presented in Sec. \ref{sec: summary}.

\section{Gravitational lensing basics}\label{sec: gravitational lensing basics}
In this section we describe some aspects of the gravitational lensing effect within the framework of the thin lens approximation. In particular, we focus on spherically symmetric mass distributions, which generate axisymmetric lenses. With respect to the general formalism for axisymmetric gravitational lenses, we mainly follow \cite{schneider1992gl,schneider2006gl} unless otherwise stated. An alternative approach to lenses with this symmetry can be found in, e.g., \cite{hurtado2014}.

Within the thin lens approximation, the lens is characterised by the projected mass distribution along the line of sight (denoted by the coordinate $z$), whose surface mass density can be written as
\begin{equation}\label{eq: surface mass density}
\Sigma(\xi)= 2 \int_{0}^{\infty}\rho(\xi,z)dz,
\end{equation}
where $\rho(r)$ is the spherically symmetric mass distribution, with $r^2 = \xi^2+z^{2}$, and $\xi$ represents the radial coordinate on the lens plane (perpendicular to line of sight). Introducing the dimensionless radius $x=\xi/\xi_0$, where $\xi_0$ serves as an adequate normalisation length scale, the projected mass enclosed by the lens within a radius $x$ (equivalent to the mass enclosed by an infinite cylinder of such radius) takes the form
\begin{equation}\label{eq: projected mass}
M(x)=2\pi \xi_0^2\int_{0}^{x}\chi\Sigma(\chi)d\chi.   
\end{equation}

Once we have a lens with redshift $z_l$, the bending of light is ruled by the lens equation, a two-dimensional vector equation that for a given source with redshift $z_s>z_l$, links its true position $\boldsymbol{\eta}$ on the source plane to its apparent position $\boldsymbol{\xi}$ on the lens plane. The corresponding deviation is given by the deflection angle $\hat{\boldsymbol{\alpha}}$. Axisymmetric lenses satisfy the condition that $\boldsymbol{\eta}$, $\boldsymbol{\xi}$, and $\hat{\boldsymbol{\alpha}}$ are collinear, allowing us to drop the azimuthal dependence and keep only their radial component, leading to the one-dimensional lens equation
\begin{equation}\label{eq: deimensionless lens equation 1d}
y=x-\alpha(x),
\end{equation}
where $y=\eta/\eta_0$ with $\eta_0=(D_s/D_l)\xi_0$, and
\begin{equation}\label{eq: deflection angle}
\alpha(x) = \dfrac{D_l D_{ls}}{D_s \xi_0}\hat{\alpha}(x) = \dfrac{2}{\Sigma_{\text{cr}}x}\int_{0}^{x} \chi\Sigma(\chi)d\chi=\dfrac{M(|x|)}{\pi \xi_{0}^{2}\Sigma_{\text{cr}}x}
\end{equation}
is the radial component of the scaled deflection angle, which we will refer to as the deflection angle for simplicity. Note that
\begin{equation}\label{eq: critical density}
\Sigma_{\text{cr}}= \frac{c^2 D_s}{4\pi G D_lD_{ls}}
\end{equation}
is the critical (lens) density. The gravitational lensing effect is mediated by the angular diameter distances  $D_s =D(z_s)$, $D_l=D(z_l)$, and $D_{ls}=D(z_l,z_s)$, which correspond to the distances between the observer and source,  observer and lens, and lens and source, respectively.

The lens equation \eqref{eq: deimensionless lens equation 1d} is non-linear in $x$, meaning that for a given source position $y$, multiple solutions $x$ may exist, each corresponding to a distinct apparent image. The deflection angle \eqref{eq: deflection angle} is expressed in terms of $M(|x|)$ to account for possible negative solutions. Beyond producing multiple images, the lens also distorts in both shape and size the apparent contour of background sources. The isotropic deformations (scaling), and anisotropic deformations (change in the apparent ellipticity) are described by the convergence
\begin{equation}\label{eq: convergence}
\kappa(x)=\dfrac{\Sigma(x)}{\Sigma_{\text{cr}}}, 
\end{equation}
and shear, respectively. For axisymmetric lenses the shear is given by \citep{miralda-escude1991}
\begin{equation}\label{eq: shear}
\gamma(x) = \dfrac{\widebar{\Sigma}(x)-\Sigma(x)}{\Sigma_{\text{cr}}},    
\end{equation}
with
\begin{equation}\label{eq: mean surface mass density}
\widebar{\Sigma}(x)=\dfrac{2}{x^2}\int_{0}^{x}\chi\Sigma(\chi)d\chi=\dfrac{M(x)}{\pi \xi_0^2x^2}
\end{equation}
being the mean surface mass density within $x$. For axisymmetric lenses, \eqref{eq: shear} equals the average tangential shear taken over a circle of radius $x$. Some applications of the tangential shear con be found in, e.g., \cite{bartelmann1995clusters_weak_lensing, nakajima2009tangential_shear, giocoli2014tangentia_shear}. A stronger signal is obtained after taking the mean value of \eqref{eq: shear} up to a given radius $x$, which leads to the mean shear
\begin{equation}\label{eq: mean shear}
\widebar{\gamma}(x)=\dfrac{2}{x^2}\int_{0}^{x}\chi\gamma(\chi)d\chi.
\end{equation}

The  gravitational lensing effect can also be described through a scalar potential known as the deflection potential, which for axisymmetric lenses takes the form 
\begin{equation}\label{eq: deflection potential}
 \psi(x)=\frac{2}{\Sigma_{\text{cr}}}\int_{0}^{x} \chi\Sigma(\chi)\ln\left(\frac{x}{\chi}\right)d\chi.
 \end{equation}
This potential relates to the deflection angle and convergence through 
\begin{equation}\label{eq: deflection potential relations}
\alpha(x)=\psi^{\prime}(x)\quad \text{and}\quad 2\kappa(x) = \dfrac{\psi^{\prime}(x)}{x}+\psi^{\prime\prime}(x), 
\end{equation}
where the prime indicates derivatives with respect to $x$. 

Now, for axisymmetric lenses we derive an explicit relation between the mean shear and the deflection potential. From \eqref{eq: shear}, the mean shear \eqref{eq: mean shear} can be written as $\overline{\gamma}(x)=(\widetilde{\Sigma}(x)-\overline{\Sigma}(x))/\Sigma_{\text{cr}}$, where $\overline{\Sigma}(x)$ is given by \eqref{eq: mean surface mass density}, and 
\begin{equation}\label{eq: sigma tilde}
\widetilde{\Sigma}(x) = \dfrac{2}{x^2}\int^{x}_0 \chi\overline{\Sigma}(\chi)d\chi=\dfrac{2}{\pi \xi_0^2 x^2}\int^{x}_0\dfrac{M(\chi)}{\chi}d\chi.
\end{equation}
Using \eqref{eq: deflection angle} and \eqref{eq: deflection potential relations}, and considering that the integration in \eqref{eq: sigma tilde} is carried out over a non-negative domain, the integrand can be expressed in terms of the deflection potential as 
\begin{equation}
\dfrac{M(\chi)}{\chi}=\pi\xi_0^2\Sigma_{\text{cr}}\psi^{\prime}(\chi).    
\end{equation}
Thus, after substituting this result into \eqref{eq: sigma tilde}, the mean shear becomes
\begin{equation}\label{eq: mean shear - potential}
\overline{\gamma}(x)=\dfrac{2\Delta\psi(x)}{x^2}-\dfrac{M(x)}{\pi \xi_0^2\Sigma_{\text{cr}}x^2}=\dfrac{2\Delta\psi(x)}{x^2}-\dfrac{\psi^{\prime}(x)}{x},
\end{equation}
with $\Delta \psi(x)=\psi(x)-\psi(0)$. Note that for \eqref{eq: mean shear - potential} to remain finite,  $\psi(0)$ must to be well defined. Expression \eqref{eq: mean shear - potential} has been used, for example, in \cite{mertel2003cdm_halos}.

\section{A general lens via the Fox $H$-function}\label{sec: general result}
Let us consider a general lens whose surface mass density can be written in terms of the Fox $H$-function as
\begin{equation}\label{eq: general surface mass density}
\Sigma(x) = \Sigma_0 x^a
\underbrace{\left(\dfrac{1}{2\pi i}
\int_{\mathcal{L}}\Theta_{\Sigma}(v)x^{-b v}dv\right)}_{\text{\normalsize Fox $H$-function, see \eqref{eq: Fox H-function}}},
\end{equation}
where $a,b\in\mathbb{R}$ with $b\neq 0$, $\Theta_{\Sigma}(v)$ satisfies \eqref{eq: Fox H-function argument}, and $\Sigma_0$ is a constant that depends on the physical properties of the lens, known as the characteristic surface mass density. 

Here, we show that relation \eqref{eq: mean shear - potential} holds true for lenses that satisfy \eqref{eq: general surface mass density} (including the gNFW), without requiring explicit knowledge of $\psi(0)$. To do this, we evaluate \eqref{eq: sigma tilde} and assess whether it can be expressed in terms of the deflection potential. First of all, we need the projected mass, which can be obtained by substituting \eqref{eq: general surface mass density} into \eqref{eq: projected mass}, such that 
\begin{equation}\label{eq: general projected mass 0}
M(x) = 2\pi\xi_0^2\Sigma_0
\left(\dfrac{1}{2\pi i}
\int_{\mathcal{L}}\int_{0}^x\Theta_{\Sigma}(v)\chi^{1+a-bv}d\chi dv\right),
\end{equation}
which, once we perform the integration with respect to $\chi$ becomes
\begin{equation}\label{eq: general projected mass}
M(x) = \dfrac{2\pi\xi_0^2\Sigma_0x^{2+a}}{b}
\left(\dfrac{1}{2\pi i}
\int_{\mathcal{L}}\Theta_{M}(v)x^{-bv}dv\right),
\end{equation}
with
\begin{equation}\label{eq: general theta projected mass}
\Theta_{M}(v) = \dfrac{1}{\frac{2+a}{b}-v}\Theta_{\Sigma}(v)= \dfrac{\Gamma\left(\frac{2+a}{b}-v\right)}{\Gamma\left(\frac{2+a+b}{b}-v\right)}\Theta_{\Sigma}(v).
\end{equation}
We work with $\left(\frac{2+a}{b}-v\right)$ instead of $\left(2+a-bv\right)$ for convenience. Now, by substituting \eqref{eq: general projected mass} into \eqref{eq: sigma tilde} and following the same procedure described above, we obtain
\begin{equation}\label{eq: general mean surface mass density}
\widetilde{\Sigma}(x) = \dfrac{4\Sigma_0 x^a}{b^2}
\left(\dfrac{1}{2\pi i}
\int_{\mathcal{L}}\Theta_{\widetilde{\Sigma}}(v)x^{-bv}dv\right),
\end{equation}
with
\begin{equation}\label{eq: general theta M}
\Theta_{\widetilde{\Sigma}}(v) = \dfrac{1}{\frac{2+a}{b}-v}\Theta_{M}(v)= \dfrac{\Gamma\left(\frac{2+a}{b}-v\right)^2}{\Gamma\left(\frac{2+a+b}{b}-v\right)^2}\Theta_{\Sigma}(v).
\end{equation}
Similarly, if we substitute \eqref{eq: general mean surface mass density} into \eqref{eq: deflection potential} and integrate with respect to $\chi$ (analogous to what we did to get \eqref{eq: general projected mass}), the deflection potential takes the form 
\begin{equation}\label{eq: general deflection potential}
\psi(x) 
= \dfrac{2\kappa_0 x^{2+a}}{b^2}
\left(\dfrac{1}{2\pi i}
\int_{\mathcal{L}}\Theta_{\psi}(v)x^{-bv}dv\right),
\end{equation}
where $\kappa_0 = \Sigma_0/\Sigma_{\text{cr}}$ is the characteristic convergence, and
\begin{equation}\label{eq: general theta deflection potential}
\Theta_{\psi}(v) = \dfrac{1}{\left(\frac{2+a}{b}-v\right)^2}\Theta_{\Sigma}(v)= \dfrac{\Gamma\left(\frac{2+a}{b}-v\right)^2}{\Gamma\left(\frac{2+a+b}{b}-v\right)^2}\Theta_{\Sigma}(v).
\end{equation}
Then, from comparing \eqref{eq: general mean surface mass density} and \eqref{eq: general deflection potential} it is clear that 
\begin{equation}
\widetilde{\Sigma}(x)=2\Sigma_{\text{cr}}\psi(x)/x^2.
\end{equation}
This result implies that for any lens with surface mass density described by \eqref{eq: general surface mass density}, its deflection potential satisfies $\psi(0)=0$, which is equivalent to have $\Delta\psi(x)=\psi(x)$. As outlined in Sec. \ref{sec: gravitational lensing basics}, it is clear that, from the surface mass density \eqref{eq: general surface mass density}, projected mass \eqref{eq: general projected mass}, and deflection potential \eqref{eq: general deflection potential}, we have all we need to describe our general lens.

Now, we evaluate the total mass $M_{\infty}=M(x\to\infty)$ enclosed by the lens. To do so, we use \eqref{eq: general projected mass 0}, where we apply the change of variable $w=\chi^{b}$ and take $x\to\infty$. If we consider $b>0$, the total mass can be written as
\begin{equation}\label{eq:  general  total mass}
M_{\infty} = 
\int_{0}^{\infty}m(w)w^{(2+a-b)/b}dw,
\end{equation}
where
\begin{equation}
m(w)=\dfrac{1}{2\pi i}
\int_{\mathcal{L}}\mathcal{M}(v)w^{-v}dv=\mathfrak{M}^{-1}_{\mathcal{M}}(w)
\end{equation}
turns out to be the inverse Mellin transform (see \eqref{eq: mellin inverse transform}) of
\begin{equation}\label{eq: general M prime}
\mathcal{M}(v)=2\pi \xi_0^2\Sigma_0\Theta_{\Sigma}(v)/b, 
\end{equation}
which, in turn, is equal to the Mellin transform (see \eqref{eq: mellin transform}) of $m(w)$. This is,
\begin{equation}\label{eq: mellin transform m}
\mathfrak{M}_{m}(v) = \int_{0}^{\infty} m(w) w^{v-1}dw= \mathcal{M}(v).   
\end{equation}
By comparing \eqref{eq: general total mass} and \eqref{eq: mellin transform m}, we obtain a simple expression for $M_{\infty}$ in terms of \eqref{eq: general M prime}. We then perform an analogous process for $b<0$ and combine both results into
\begin{equation}\label{eq: general total mass final}
M_{\infty}=
\begin{dcases}
\mathcal{M}\left(\frac{2+a}{b}\right)   & \text{if } b>0\\[1mm]
-\mathcal{M}\left(\frac{2+a}{b}\right)   & \text{if } b<0
\end{dcases}.
\end{equation}


\section{The generalised-NFW halo as a gravitational lens}\label{sec: gNFW lensing}
If we want to apply to the gNFW the framework presented in the previous section, we need to show first that its surface mass density satisfies \eqref{eq: general surface mass density}. To do this, we follow the methodology outlined in Appendix \ref{appendix: mellin}. We proceed by expressing the surface mass density as the Mellin convolution $f_1\star f_2$ (as defined in \eqref{eq: mellin convolution}) between two suitable functions, $f_1$ and $f_2$. The definition of the surface mass density given in \eqref{eq: surface mass density} is not suited for the job, instead, it can be rewritten as the Abel integral
\begin{equation}\label{eq: surface mass density abel}
\Sigma(\xi) = 2\int_\xi^{\infty} \dfrac{\rho(r)rdr}{\sqrt{r^2-\xi^2}}, 
\end{equation}
which,  for the gNFW with normalisation length $\xi_0=r_0$ and dimensionless radius $x=\xi/r_0$, reads
\begin{equation}\label{eq: gnfw surface mass density abel}
\Sigma(x) = \dfrac{(n-2)\phi_{0}}{2\pi G r_0}\int_x^{\infty} \dfrac{dt}{(1+t)^{n-1}\sqrt{t^{2}-x^2}}, 
\end{equation}
where $t=r/r_0$. For \eqref{eq: gnfw surface mass density abel} the required functions $f_1$ and $f_2$ take the form
\begin{equation}\label{eq: gNFW f1}
f_1(h=t) = \dfrac{(n-2)\phi_{0}}{2\pi G r_0}\dfrac{h}{(1+h)^{n-1}},
\end{equation}
and
\begin{equation}\label{eq: gNFW f2}
f_2(h=x/t) = 
\begin{cases}
 \dfrac{h}{x\sqrt{1-h^2}} & \text{if}\quad 0\leq h < 1\\
0 & \text{otherwise}
\end{cases}.
\end{equation}
The next step requires the Mellin transform $\mathfrak{M}$ (as defined in \eqref{eq: mellin transform}) of $f_1$ and $f_2$, which are
\begin{equation}\label{eq: gNFW mellin transform 1}
\mathfrak{M}_{f_1}(u)= \dfrac{(n-2)\psi_{0}}{2\pi G a}\dfrac{\Gamma\left(n-2-u\right)\Gamma(1+u)}{\Gamma(n-1)}
\end{equation}
and
\begin{equation}\label{eq: gNFW mellin transform 2}
\mathfrak{M}_{f_2}(u)= \dfrac{\sqrt{\pi}}{2x}\dfrac{\Gamma\left(\frac{1+u}{2}\right)}{\Gamma\left(1+\frac{u}{2}\right)},
\end{equation}
respectively. Since the surface mass density satisfies $\Sigma(x)=(f_1\star f_2)(x)$, it can also be expressed following the definition in \eqref{eq: mellin transform technique}. Therefore, we substitute \eqref{eq: gNFW mellin transform 1} and \eqref{eq: gNFW mellin transform 2} into \eqref{eq: mellin transform technique}, and (for convenience) apply the change of variable $v=1+u/2$, leading to
\begin{equation}\label{eq: gNFW surface mass density Mellin}
\Sigma(x) = \Sigma_{0} x
\left(\dfrac{1}{2\pi i}
\int_{\mathcal{L}}\Theta_{\Sigma}(v)x^{-2v}dv\right),
\end{equation}
with
\begin{equation}\label{eq: gnfw theta surface mass density}
\Theta_{\Sigma}(v)= \dfrac{\Gamma\left(-1+2v\right) \Gamma\left(-\frac{1}{2}+v\right)\Gamma\left(n-2v\right)}{\Gamma(v)}
\end{equation}
and
\begin{equation}\label{eq: gnfw surface mass density 0}
\Sigma_{0}=\dfrac{(n-2)\phi_0}{2\sqrt{\pi} G r_0\Gamma(n-1)}.
\end{equation}
We can see that, with $a=1$ and $b=2$, the gNFW falls within the family of lenses defined by \eqref{eq: general surface mass density}.

As outlined in Sec. \ref{sec: general result}, from \eqref{eq: gnfw theta surface mass density} each $\Theta_{i}(v)$ (with $i=M$ \eqref{eq: general theta projected mass}, $\psi$ \eqref{eq: general theta deflection potential}) takes the form
\begin{equation}\label{eq: gnfw theta projected mass}
\Theta_{M}(v)= \dfrac{\Gamma\left(-1+2v\right) \Gamma\left(-\frac{1}{2}+v\right)\Gamma\left(n-2v\right)\Gamma\left(\frac{3}{2}-v\right)}{\Gamma(v)\Gamma\left(\frac{5}{2}-v\right)},
\end{equation}
\begin{equation}\label{eq: gnfw theta deflection potential}
\Theta_{\psi}(v)= \dfrac{\Gamma\left(-1+2v\right) \Gamma\left(-\frac{1}{2}+v\right)\Gamma\left(n-2v\right)\Gamma\left(\frac{3}{2}-v\right)^2}{\Gamma(v)\Gamma\left(\frac{5}{2}-v\right)^2}.
\end{equation}
We apply these results to the definition of the Fox $H$-function given in \eqref{eq: Fox H-function}, obtaining
\begin{equation}\label{eq: gNFW surface mass density H}
\displaystyle
\Sigma(x)=\Sigma_0 x H_{2,2}^{2,1} \!\left[ x^2 \left| \begin{matrix}
( 1-n , 2 ),( 0 , 1 ) \\
( -1 , 2 ),( -\frac{1}{2} , 1 ) \end{matrix} \right. \right],
\end{equation}
\begin{equation}\label{eq: gNFW projected mass H}
\displaystyle
M(x)=\pi r_0^2\Sigma_0 x^3 H^{2,2}_{3,3} \!\left[ x^2 \left| \begin{matrix}
(1-n,2), ( -\frac{1}{2} , 1 ), (0, 1) \\
( -1 , 2 ),( -\frac{1}{2} , 1), (-\frac{3}{2}, 1) \end{matrix} \right. \right],
\end{equation}
\begin{equation}\label{eq: gNFW deflection potential H}
\displaystyle
\psi(x)=\frac{\kappa_0 x^3}{2}
 H^{2,3}_{4,4} \!\left[ x^2 \left| \begin{matrix}
(1-n,2),  ( -\frac{1}{2} , 1 ),  ( -\frac{1}{2} , 1 ), (0, 1) \\
( -1 , 2 ), ( -\frac{1}{2} , 1 ),  ( -\frac{3}{2} , 1 ),  ( -\frac{3}{2} , 1)  \end{matrix} \right. \right].
\end{equation}
Each of these Fox $H$-functions can be written in its short notation as 
\begin{equation}\label{eq: fox h short}
H^{2,1}_{2,2\,\Sigma}\left(x^2,n\right),\; H^{2,2}_{3,3\,M}\left(x^2,n\right),\; H^{2,3}_{4,4\,\psi}\left(x^2,n\right),  
\end{equation}
respectively. Thus, the convergence \eqref{eq: convergence} and deflection angle \eqref{eq: deflection angle} become
\begin{equation}\label{eq: gnfw convergence H}
 \kappa(x) = \kappa_0 x H^{2,1}_{2,2\,\Sigma}\left(x^2,n\right)
\end{equation}
and
\begin{equation}\label{eq: gnfw deflection angle H}
\alpha(x) = \dfrac{\kappa_0|x|^3}{x} H^{2,2}_{3,3\,M}\left(x^2,n\right),
\end{equation}
while the shear \eqref{eq: shear} and mean shear \eqref{eq: mean shear - potential} are given by
\begin{equation}\label{eq: gnfw shear H}
\gamma(x) = \kappa_0x\left( H^{2,2}_{3,3\,M}\left(x^2,n\right)-H^{2,1}_{2,2\,\Sigma}\left(x^2,n\right)\right)
\end{equation}
and
\begin{equation}\label{eq: gnfw mean shear H}
\overline{\gamma}(x) = \kappa_0x\left( H^{2,3}_{4,4\,\psi}\left(x^2,n\right) - H^{2,2}_{3,3\,M}\left(x^2,n\right)\right).
\end{equation}

From either the power series listed in Sec. \ref{sec: series representation} or the expression in Sec. \ref{sec: final expressions}, it can be shown that, in the limit as $x\to 0$, the surface mass density and convergence diverge. The projected mass, deflection angle, and deflection potential satisfy $M(x\to 0)=\alpha(x\to 0)=\psi(x\to 0)=0$, and for the shear and mean shear we get  
\begin{equation}\label{eq: shear x=0}
\gamma(x\to 0)=\overline{\gamma}(x\to 0)=\dfrac{\kappa_0\Gamma(n-1)}{2\sqrt{\pi}}.   
\end{equation}

Additionally, substituting \eqref{eq: gnfw theta surface mass density} and \eqref{eq: gnfw surface mass density 0} into \eqref{eq: general total mass final} yields the total mass enclosed by the gNFW lens, which turns out to be 
\begin{equation}\label{eq: gnfw total mass}
M_{\infty}=\mathcal{M}(3/2)=\dfrac{r_0(n-2)\Gamma(n-3)\phi_0}{G\Gamma(n-1)}=\dfrac{r_0\phi_0}{G(n-3)}.
\end{equation}
We can see that $M_{\infty}$ diverges or is negative for $2<n\leq 3$, but remains finite and positive for $n>3$. This result is in agreement with the total mass reported by \cite{evansGNFW2006}.

As a final remark, it is worth noting that there are exceptions to the rule, and for some mass distributions (with spherical symmetry) the Mellin technique simply does not work. For instance, that is the case with the Singular Isothermal Sphere (SIS). We elaborate a little bit more about this limitation in Appendix \ref{appendix: sis and nis}.

\subsection{Meijer $G$-function representation}
The Fox $H$-function has not yet been implemented in most software used for numerical calculations. That is not the case of the Meijer G-function, which is implemented, for example, in \textsc{mathematica} and \textsc{python}. Depending on the application in mind or the software being used, these implementations (on their current state) may be impractical. Nonetheless, the representation of \eqref{eq: gNFW surface mass density H}-\eqref{eq: gNFW deflection potential H} in terms of the Meijer $G$-function might be handy. Since the Meijer $G$-function is a special case of the Fox $H$-function (see Appendix \ref{appendix: fox}), from Gausss multiplication formula \citep{abramowitz1972hmfw}
\begin{equation}
\displaystyle
\Gamma(kx)=(2\pi)^{(1-k)/2}k^{kx-1/2}\prod_{j=0}^{k-1}\Gamma\left(x+\frac{j}{k}\right),    
\end{equation}
where $k\in\mathbb{Z}^{+}$, we can rewrite \eqref{eq: gnfw theta surface mass density} as
\begin{equation}
\Theta_{\Sigma}(v) = \dfrac{2^{n-3}}{\pi}\Gamma\left(-\frac{1}{2}+v\right)^2 \Gamma\left(\frac{n+1}{2}-v\right)\Gamma\left(\frac{n}{2}-v\right),  
\end{equation}
which allows us to write  \eqref{eq: gNFW surface mass density H}-\eqref{eq: gNFW deflection potential H} as
\begin{equation}\label{eq: gNFW surface mass density G}
\displaystyle
\Sigma(x)=\frac{2^{n-3}\Sigma_0 x}{\pi} G^{2,2}_{2,2} \!\left[ x^2 \left| \begin{matrix}
\frac{1-n}{2}, 1 - \frac{n}{2} \\
-\frac{1}{2}, -\frac{1}{2} \end{matrix} \right. \right],
\end{equation}
\begin{equation}\label{eq: gNFW projected mass G}
\displaystyle
M(x)=2^{n-3}r_0^2\Sigma_0 x^3 G^{2,3}_{3,3} \!\left[ x^2 \left| \begin{matrix}
\frac{1-n}{2}, 1 - \frac{n}{2}, -\frac{1}{2} \\
-\frac{1}{2}, -\frac{1}{2}, -\frac{3}{2} \end{matrix} \right. \right],
\end{equation}
\begin{equation}\label{eq: gNFW deflection potential G}
\displaystyle
\psi(x)=\frac{2^{n-4}\kappa_0 x^3}{\pi} G^{2,4}_{4,4} \!\left[ x^2 \left| \begin{matrix}
\frac{1-n}{2}, 1 - \frac{n}{2}, -\frac{1}{2}, -\frac{1}{2} \\
-\frac{1}{2}, -\frac{1}{2}, -\frac{3}{2}, -\frac{3}{2}  \end{matrix} \right. \right].
\end{equation}


\subsection{Power series representation}\label{sec: series representation}
From what we have discussed so far, we require three independent  representations of the Fox $H$-function (given in \eqref{eq: fox h short}). The domain in which these functions converge depends on the parameters $\Delta$ and $\delta$ (defined in \eqref{eq: convergence conditions H}), as detailed in Appendix \ref{cases}. In our case, all three functions satisfy $\Delta=0$ and $\delta=1$, such that they are well-defined for $0<x^2<1$ and $x^2>1$. 

Regarding their power series representation, for each domain we need to evaluate the poles of the different $\Gamma$ functions present in the numerator of each $\Theta_{i}(v)$ (with $i=\Sigma$ \eqref{eq: gnfw theta surface mass density}, $M$ \eqref{eq: gnfw theta projected mass}, $\psi$ \eqref{eq: gnfw theta deflection potential}). For $0<x^2<1$ we consider functions of the form $\Gamma(b_j+B_j v)$, whose poles satisfy \eqref{eq: poles beta}. Similarly, for $x^2>1$ we consider (the remaining) functions of the form $\Gamma(1-a_j-A_j v)$, whose poles satisfy \eqref{eq: poles alpha}. The general expansions considering poles up to third order are covered in \eqref{eq: H series general}. In general, they depend not only  on $\Gamma$ functions, but also on the special functions $\psi_0$ and $\psi_1$, known as the digamma and trigamma functions, respectively. Note that for \eqref{eq: H series general} to apply, there must be no overlap between the poles \eqref{eq: poles beta} and the poles \eqref{eq: poles alpha}. See Appendix \ref{appendix: fox} for further details on the Fox $H$-function and its power series representations.

\subsubsection{Function $H^{2,1}_{2,2\,\Sigma}\left(x^2,n\right)$}
From \eqref{eq: gnfw theta surface mass density}, we have that within the domain $0<x^2<1$, the poles are
\begin{align}\label{eq: poles sigma b}
\nonumber
\Gamma(b_j+B_j v)\quad\longrightarrow\quad
&\beta_{1} = (1-k_1)/2\\
&\beta_{2} = 1/2-k_{2},
\end{align}
with $k_1,k_2\in\mathbb{N}_0$ (keep this in mind). We have poles of second order when the constraint $\beta_1=\beta_2$ is satisfied, which is equivalent to have $k_1=2k_2$. This constraint indicates that poles $\beta_1$ with $k_1$ even, and all poles $\beta_2$ are of second order. In consequence, poles $\beta_1$ with $k_1$ odd are simple. For $x^2>1$ we only have single poles of the form
\begin{align}
\Gamma(1-a_j-a_j v)\quad\longrightarrow\quad
\alpha = (n+l)/2,
\end{align}
with $l\in\mathbb{N}_0$.

The corresponding expansions on each domain are given below.

\vspace{2 mm}
\noindent
\textbf{For $0<x^2<1$ ($n>2$):}
\begin{multline}\label{eq: series smd 0<x<1}
\frac{2^{n-3}}{\pi|x|}\sum_{k=0}^{\infty}\Gamma\left(\frac{n-1}{2}+k\right)\Gamma\left(\frac{n}{2}+k\right)\dfrac{x^{2k}}{(k!)^2}\Bigg[2\psi_{0}(1+k)\\
-\psi_{0}\left(\dfrac{n-1}{2}+k\right)-\psi_{0}\left(\dfrac{n}{2}+k\right)-\ln\left(x^2\right)\Bigg]
\end{multline}

\noindent
\textbf{For $x^2>1$  ($n>2$):}
\begin{align}
&\frac{1}{2}\sum_{l=0}^{\infty}\dfrac{(-1)^{l}}{l!}\dfrac{\Gamma\left(n-1+l\right)\Gamma\left(\frac{n-1+l}{2}\right)}{\Gamma\left(\frac{n+l}{2}\right)}|x|^{-(l+n)}
\end{align}

\subsubsection{Function $H^{2,2}_{3,3\,M}\left(x^2,n\right)$}
From \eqref{eq: gnfw theta projected mass}, we have that for $0<x^2<1$, the required poles are identical to those in \eqref{eq: poles sigma b} and therefore satisfy the same multiplicity conditions. While for $x^2>1$ the poles are
\begin{align}\label{eq: poles M a}
\nonumber
\Gamma(1-a_j-a_j v)\quad\longrightarrow\quad
&\alpha_1 = (n+l_1)/2\\
&\alpha_2 = 3/2+l_2,
\end{align}
with $l_1,l_2\in\mathbb{N}_0$ (keep this in mind). We have poles of second order when the constraint $\alpha_1=\alpha_2$ is satisfied, which equivalent to have
\begin{equation}\label{eq: constrait l pm}
l_1 = 3-n+2l_2\quad\text{or}\quad l_2=(n-3+l_1)/2.
\end{equation}
On the one hand, if $n\notin\mathbb{Z}$, from constraint \eqref{eq: constrait l pm} we have that all poles $\alpha_1$ and $\alpha_2$ are simple, since for a given $l_2$ the condition $l_1\in\mathbb{N}_0$ is not satisfied. On the other hand, if $n\in\mathbb{Z}$, poles $\alpha_1$ with $l_1$ satisfying $(n-3+l_1)\,mod\,2 = 0$ (where $l_1$ must be odd when $n$ is even, and even when $n$ is odd) are of second order, which is necessary for condition $l_2\in\mathbb{N}_0$ to be satisfied in \eqref{eq: constrait l pm}. Similarly, poles $\alpha_2$ 
with  $l_2\geq(n-3)/2$ are of second order, which is the result of considering that $l_1\geq 0$ in \eqref{eq: constrait l pm}. 

Consequently, poles $\alpha_1$ are simple for any $l_1$ satisfying $(n-3+l_1)\,mod\,2 \neq 0$ (where $l_1$ must be odd when $n$ is odd, and even when $n$ is even), as do poles $\alpha_2$ with $0\leq l_2<(n-3)/2$. For instance, if we consider $n=3$, the conditions for getting poles of second order become $l_1\,mod\,2 = 0$ and $l_2\geq 0$. This implies that poles $\alpha_1$ with $l_1$ even, and all poles $\alpha_2$ are of second order. While poles $\alpha_1$ with $l_1$ odd are simple.  

The corresponding expansions on each domain are given below.

\vspace{2 mm}
\noindent
\textbf{For $0<x^2<1$ ($n>2$):}
\begin{multline}\label{eq: series pm 0<x<1}
\frac{2^{n-3}}{\pi|x|}\sum_{k=0}^{\infty}\Gamma\left(\frac{n-1}{2}+k\right)\Gamma\left(\frac{n}{2}+k\right)\dfrac{x^{2k}}{k!(k+1)!}\Bigg[\psi_{0}(1+k)\\
+\psi_{0}(2+k)-\psi_{0}\left(\dfrac{n-1}{2}+k\right)-\psi_{0}\left(\dfrac{n}{2}+k\right)-\ln\left(x^2\right)\Bigg]
\end{multline}

\noindent
\textbf{For $x^2>1$ ($n>2$ and $n\notin \mathbb{Z}$):}
\begin{align}
\sum_{l=0}^{\infty}\dfrac{(-1)^{l}}{l!}\dfrac{\Gamma\left(n-1+l\right)\Gamma\left(\frac{n-1+l}{2}\right)}{\Gamma\left(\frac{n+l}{2}\right)}\frac{|x|^{-(l+n)}}{(3-n-l)}+\dfrac{2\Gamma(n-3)}{\sqrt{\pi}|x|^3}
\end{align}

\noindent
\textbf{For $x^2>1$ ($n=3$):}
\begin{align}\label{eq: series pm n=3 x>1}
\sum_{l=0}^{\infty}\dfrac{\Gamma\left(\frac{1}{2}+l\right)}{\Gamma(1+l)}|x|^{-(2l+4)} - \sum_{l=1}^{\infty}\dfrac{\Gamma(l)}{\Gamma\left(\frac{1}{2}+l\right)}|x|^{-(2l+3)} +\frac{2\ln\left(\frac{|x|}{2}\right)}{\sqrt{\pi}|x|^3}
\end{align}

\noindent
\textbf{For $x^2>1$ ($n>3$ and even):}
\begin{multline}
\sum_{l=0}^{\infty}\dfrac{1}{(2l)!}\dfrac{\Gamma\left(n-1+2l\right)\Gamma\left(\frac{n-1+2l}{2}\right)}{\Gamma\left(\frac{n+2l}{2}\right)}\frac{|x|^{-(2l+n)}}{(3-n-2l)}\\
+\sum_{l=(n-2)/2}^{\infty}\dfrac{\Gamma(1+2l)\Gamma(l)}{\Gamma\left(\frac{1}{2}+l\right)}\frac{|x|^{-(2l+3)}}{(3-n+2l)!}+ \dfrac{2\Gamma(n-3)}{\sqrt{\pi}|x|^3}
\end{multline}

\noindent
\textbf{For $x^2>1$ ($n>3$ and odd):}
\begin{multline}
-\sum_{l=0}^{\infty}\dfrac{1}{(2l+1)!}\dfrac{\Gamma\left(n+2l\right)\Gamma\left(\frac{n+2l}{2}\right)}{\Gamma\left(\frac{n+1+2l}{2}\right)}\frac{|x|^{-(2l+n+1)}}{(2-n-2l)}\\
-\sum_{l=(n-3)/2}^{\infty}\dfrac{\Gamma(1+2l)\Gamma(l)}{\Gamma\left(\frac{1}{2}+l\right)}\frac{|x|^{-(2l+3)}}{(3-n+2l)!}+ \dfrac{2\Gamma(n-3)}{\sqrt{\pi}|x|^3}
\end{multline}


\subsubsection{Function $H^{2,3}_{4,4\,\psi}\left(x^2,n\right)$}
From \eqref{eq: gnfw theta deflection potential}, we have that for $0<x^2<1$, the required poles are identical to those in \eqref{eq: poles sigma b} and therefore satisfy the same multiplicity conditions. While for $x^2>0$, the poles are
\begin{align}\label{eq: poles shear mean a}
\nonumber
\Gamma(1-a_j-a_j v)\quad\longrightarrow\quad
&\alpha_1 = (n+l_1)/2\\\nonumber
&\alpha_2 = 3/2+l_2\\
&\alpha_3 = 3/2+l_3
\end{align}
with $l_1,l_2,l_3\in\mathbb{N}_0$ (keep this in mind). Here, we can get poles up to third order, which occur when  $\alpha_1=\alpha_2=\alpha_3$. Since $\alpha_2=\alpha_3$ for all $l_2$ and $l_3$, these are always poles of at least second order. Thus, poles of third order occur when $\alpha_1 = \alpha_i$ (with $i=1,2$). The remaining poles $\alpha_1$ are simple.  Additionally, by comparing poles in \eqref{eq: poles shear mean a} and \eqref{eq: poles M a}, we can directly infer the multiplicity for poles in \eqref{eq: poles shear mean a}. 

Therefore, if $n\notin\mathbb{Z}$, all poles $\alpha_1$ are simple, while  $\alpha_2$ and $\alpha_3$ are poles of second order, since $\alpha_2=\alpha_3$. If $n\in\mathbb{Z}$,   
poles $\alpha_1$ with $l_1$ satisfying $(n-3+l_1)\,mod\,2 = 0$ (where $l_1$ must be odd when $n$ is even, and even when $n$ is odd) are of third order, as do poles $\alpha_i$ with  $l_i\geq(n-3)/2$. Consequently, poles $\alpha_1$ are simple for any $l_1$ satisfying $(n-3+l_1)\,mod\,2 \neq 0$ (where $l_1$ must be odd when $n$ is odd, and even when $n$ is even), while $\alpha_i$ are poles of second order for $0\leq l_i<(n-3)/2$, since $\alpha_2=\alpha_3$. In this case, if we consider $n=3$, all $\alpha_2$ and $\alpha_3$, and $\alpha_1$ with $l_1$ even are poles of third order, while poles $\alpha_1$ with $l_1$ odd are simple. 

The corresponding expansions on each domain are given below, where $ \gamma_0=-\psi_{0}(1)$ represents the Euler-Mascheroni constant.

\vspace{2 mm}
\noindent
\textbf{For $0<x^2<1$ ($n>2$):}
\begin{multline}\label{eq: series dp 0<x<1}
\frac{2^{n-3}}{\pi|x|}\sum_{k=0}^{\infty}\Gamma\left(\frac{n-1}{2}+k\right)\Gamma\left(\frac{n}{2}+k\right)\dfrac{x^{2k}}{((k+1)!)^2}\Bigg[2\psi_{0}(2+k)\\
-\psi_{0}\left(\dfrac{n-1}{2}+k\right)-\psi_{0}\left(\dfrac{n}{2}+k\right)-\ln\left(x^2\right)\Bigg]
\end{multline}

\noindent
\textbf{For $x^2>1$ ($n>2$ and $n\notin \mathbb{Z}$):}
\begin{multline}
2\sum_{l=0}^{\infty}\dfrac{(-1)^{l}}{l!}\dfrac{\Gamma\left(n-1+l\right)\Gamma\left(\frac{n-1+l}{2}\right)}{\Gamma\left(\frac{n+l}{2}\right)}\frac{|x|^{-(l+n)}}{(3-n-l)^2}\\
+\dfrac{4\Gamma(n-3)}{\sqrt{\pi}|x|^3}\left(\ln\left(\frac{|x|}{2}\right)+\psi_{0}(n-3)+\gamma_0\right)
\end{multline}

\noindent
\textbf{For $x^2>1$ ($n=3$):}
\begin{multline}\label{eq: series dp n=3 x>1}
-4\sum_{l=0}^{\infty}\dfrac{\Gamma\left(\frac{3}{2}+l\right)}{\Gamma(1+l)}\frac{|x|^{-(2l+4)}}{(1+2l)^2} + \sum_{l=1}^{\infty}\dfrac{\Gamma(1+l)}{\Gamma\left(\frac{1}{2}+l\right)}\frac{|x|^{-(2l+3)}}{l^2}\\
 +\frac{2}{\sqrt{\pi}|x|^3}\left(\frac{\pi}{4}^2+\ln^2\left(\frac{|x|}{2}\right)\right)
\end{multline}

\noindent
\textbf{For $x^2>1$ ($n>3$ and even):}
\begin{multline}
\sum_{l=0}^{\infty}\dfrac{2}{(2l)!}\dfrac{\Gamma\left(n-1+2l\right)\Gamma\left(\frac{n-1+2l}{2}\right)}{\Gamma\left(\frac{n+2l}{2}\right)}\frac{|x|^{-(2l+n)}}{(3-n-2l)^2}\\
-\sum_{ l=(n-2)/2}^{\infty}\dfrac{\Gamma(1+2l)\Gamma(1+l)}{\Gamma\left(\frac{1}{2}+l\right)}\frac{|x|^{-(2l+3)}}{l^2(3-n+2l)!}\\
+\dfrac{4\Gamma(n-3)}{\sqrt{\pi}|x|^3}\left(\ln\left(\frac{|x|}{2}\right)+\psi_{0}(n-3)+\gamma_0\right)
\end{multline}

\noindent
\textbf{For $x^2>1$ ($n>3$ and odd):}
\begin{multline}
-\sum_{l=0}^{\infty}\dfrac{2}{(2l+1)!}\dfrac{\Gamma\left(n+2l\right)\Gamma\left(\frac{n+2l}{2}\right)}{\Gamma\left(\frac{n+1+2l}{2}\right)}\frac{|x|^{-(2l+n+1)}}{(2-n-2l)^2}\\
+\sum_{ l=(n-3)/2}^{\infty}\dfrac{\Gamma(1+2l)\Gamma(1+l)}{l^2\Gamma\left(\frac{1}{2}+l\right)}\frac{|x|^{-(2l+3)}}{(3-n+2l)!}\\
+\dfrac{4\Gamma(n-3)}{\sqrt{\pi}|x|^3}\left(\ln\left(\frac{|x|}{2}\right)+\psi_{0}(n-3)+\gamma_0\right)
\end{multline}


\subsection{Final expressions}\label{sec: final expressions}
In this section, we show that the power series derived in Sec. \ref{sec: series representation} can be simplified and expressed in terms of Gauss hypergeometric function ${}_2F_1$ (for more details on this function see e.g., \cite{abramowitz1972hmfw}, chapter 15). We begin with the series with domain $0<x^2<1$, where the simplified expressions for \eqref{eq: series smd 0<x<1} and \eqref{eq: series dp 0<x<1} are obtained using  Eq. (15.3.10) (from \cite{abramowitz1972hmfw}), while for \eqref{eq: series pm 0<x<1} we use Eq. (15.3.11) (from \cite{abramowitz1972hmfw}). By applying Eq. (15.3.8) (from \cite{abramowitz1972hmfw}) to each of the previous Gauss hypergeometric functions, we obtain the corresponding simplified expressions for the power series in the domain $x^2>1$. Then, we use Eq. (15.8.27) and Eq. (15.8.28) (from \cite{abramowitz1972hmfw}) as appropriate, to get more compact expressions. Recall that for us $n>2$, which is a physical constraint required by \eqref{eq: gnfw density}.

Following the procedure described above, the Fox $H$-function in \eqref{eq: gNFW surface mass density H} can be expressed as
\begin{equation}\label{eq: smd h hypergeometric}
H^{2,1}_{2,2\,\Sigma}\left(x^2,n\right)=\dfrac{\Gamma\left(n-1\right)^2}{2^{n-1}\Gamma\left(n-\frac{1}{2}\right)|x|}f_0(x),
\end{equation}
with
\begin{align}\nonumber
&f_{0}(x)= \hspace*{\fill}\qquad\\
&\times
\begin{dcases} 
 {}_2F_1\left(\frac{n-1}{2},\frac{n}{2}; n -\frac{1}{2};1-x^2\right) &\text{if } 0<x^2<1\\[1mm]
1 & \text{if }x^2= 1\\[1mm]
\dfrac{1}{|x|^{n-1}} {}_2F_1\left(n-1,n-1; n - \frac{1}{2};\frac{|x|-1}{2|x|}\right)&\text{if } x^2>1
\end{dcases}.
\end{align}
In the same way, the Fox $H$-function in \eqref{eq: gNFW projected mass H} takes the form
\begin{equation}\label{eq: h pm final expression n neq 3}
H^{2,2}_{3,3\,M}\left(x^2,n\neq 3\right)=
\dfrac{2\Gamma(n-3)}{\sqrt{\pi}|x|^3} - \dfrac{\Gamma\left(n-1\right)\Gamma\left(n-3\right)}{2^{n-3}\Gamma\left(n-\frac{3}{2}\right)|x|^3}f_{1}(x),  
\end{equation}
with
\begin{align}\label{eq: f1}
\nonumber
&f_{1}(x)=\hspace*{\fill}\qquad\\
&\times
\begin{dcases} 
{}_2F_1\left(\frac{n-3}{2},\frac{n}{2}-1; n -\frac{3}{2};1-x^2\right) &\text{if } 0<x^2<1\\[1mm]
1 &\text{if } x^2= 1\\[1mm]
\dfrac{1}{|x|^{n-3}} {}_2F_1\left(n-3,n-1; n - \frac{3}{2};\frac{|x|-1}{2|x|}\right)& \text{if } x^2>1
\end{dcases},
\end{align}
when $n\neq3$, while for $n=3$ we get
\begin{align}\label{eq: h pm final expression n = 3}
\nonumber
&H^{2,2}_{3,3\,M}\left(x^2,n=3\right)=\dfrac{2}{\sqrt{\pi}|x|^3}\hspace*{\fill}\qquad\\[1.5mm]
&\times
\begin{dcases} 
\ln\left(\dfrac{|x|}{2}\right)+\dfrac{\arctanh\left(\sqrt{1-x^2}\right)}{\sqrt{1-x^2}}&\text{if } 0<x^2<1\\[1mm]
1-\ln(2) &\text{if } x^2= 1\\[1mm]
\ln\left(\dfrac{|x|}{2}\right)+\dfrac{\arctan\left(\sqrt{x^2-1}\right)}{\sqrt{x^2-1}}& \text{if } x^2>1
\end{dcases},
\end{align}
which is the result of evaluating series \eqref{eq: series pm 0<x<1} and \eqref{eq: series pm n=3 x>1}.

For the Fox $H$-function in \eqref{eq: gNFW deflection potential H} we get an indetermination not only at $n=3$, but also at $n=5/2$. We remove the indetermination at $n=5/2$  using the contiguous relation (15.2.27) (from \cite{abramowitz1972hmfw}) in the domain $0<x^2<1$. After that, we apply the procedure outlined at the beginning of this section to extend the result to the domain $x^2>1$. Thus, when $n\neq 3$ we get
\begin{align}
\nonumber
H^{2,3}_{4,4\,\psi}\left(x^2, n\neq 3\right)=&
\dfrac{4\Gamma(n-3)}{\sqrt{\pi}|x|^3}\left(\ln\left(\frac{|x|}{2}\right)+\psi_{0}(n-3)+\gamma_0\right) \\\nonumber
&+ \dfrac{\Gamma\left(n-3\right)^2}{2^{n-4}\Gamma\left(n-\frac{3}{2}\right)|x|^5}\bigg[\left((2n-3)x^2-2\right)f_{1}(x) \\
&+ \dfrac{n(n-1)(1-x^2)}{2n-3}f_2(x)\bigg],
\end{align}
where $f_1(x)$ is given by \eqref{eq: f1} and
\begin{align}\label{eq: f2}
\nonumber
&f_{2}(x)=\hspace*{\fill}\qquad\\
&\times
\begin{dcases} 
{}_2F_1\left(\frac{n-3}{2},\frac{n}{2}-1; n -\frac{1}{2};1-x^2\right) &\text{if } 0<x^2<1\\[1mm]
1 &\text{if } x^2 = 1\\[1mm]
\dfrac{1}{|x|^{n-3}} {}_2F_1\left(n-3,n+1; n - \frac{1}{2};\frac{|x|-1}{2|x|}\right)&\text{if } x^2 > 1
\end{dcases},
\end{align}
while for $n=3$ we get
\begin{align}\label{eq: h dp final expression n = 3}
\nonumber
&H^{2,3}_{4,4\,\psi}\left(x^2,n=3\right)=\dfrac{2}{\sqrt{\pi}|x|^3}\hspace*{\fill}\qquad\\[1.5mm]
&\times
\begin{dcases} 
\ln^2\left(\dfrac{|x|}{2}\right)-\arctanh^2\left(\sqrt{1-x^2}\right)&\text{if } 0<x^2<1\\[1mm]
\ln^2(2) &\text{if } x^2= 1\\[1mm]
\ln^2\left(\dfrac{|x|}{2}\right)+\arctan^2\left(\sqrt{x^2-1}\right) & \text{if } x^2>1
\end{dcases},
\end{align}
which is the result of evaluating series \eqref{eq: series dp 0<x<1} and \eqref{eq: series dp n=3 x>1}.

\begin{figure}
\centering
\includegraphics[width=0.9\linewidth]{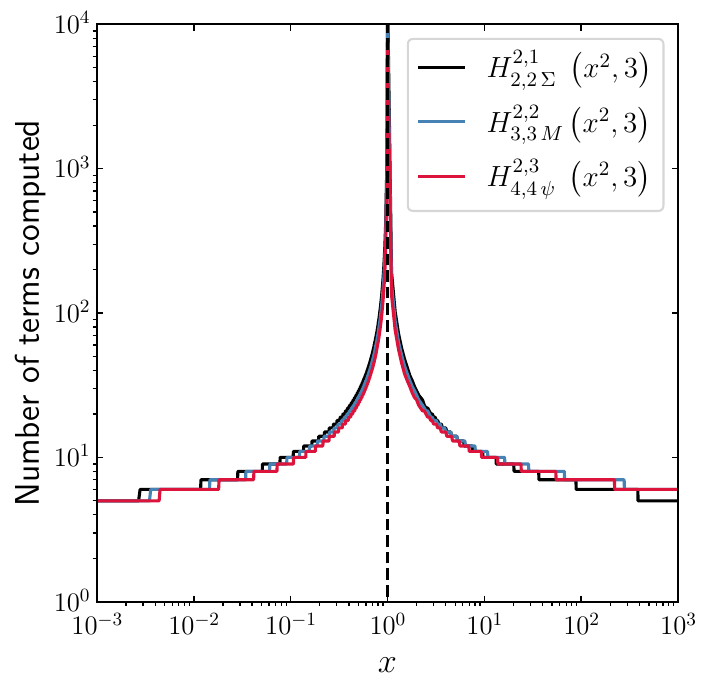}
\caption{Number of terms required for the power series representations of the relevant Fox $H$-functions to converge. In this example we consider $n=3$ and a tolerance of $10^{-15}$.}
\label{fig: N vs x}
\end{figure}
\begin{figure*}
\centering
\includegraphics[width=0.9\linewidth]{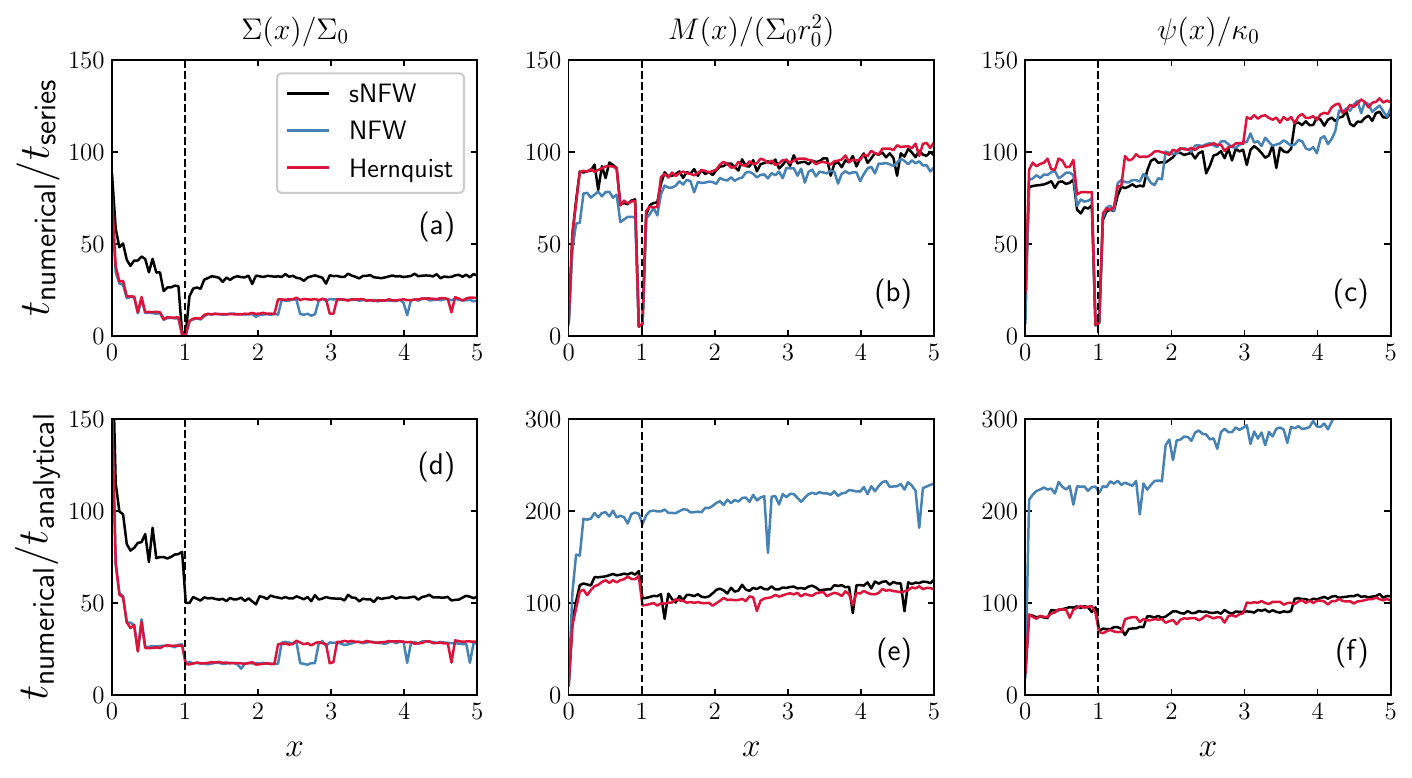}
\caption{Computation time comparison between a direct numerical integration ($t_{\text{numerical}}$) and the power series approach ($t_{\text{series}}$) (upper row), and also between the numerical integration and the analytical expressions in Sec. \ref{sec: final expressions} (lower row). The comparisons are depicted for the normalised surface mass density (left-hand column), normalised projected mass (middle column), and normalised deflection potential (right-hand column). In all panels the solid curves correspond to the sNFW ($n=7/2$, black), NFW ($n=3$, blue), and Hernquist ($n=4$, red) lenses. The vertical dashed lines represent $x=1$. The reported computation times represent averages over multiple realizations of each calculation to minimize numerical noise.}
\label{fig: computation time performance}
\end{figure*}
\begin{figure*}
\centering
\includegraphics[width=0.91\linewidth]{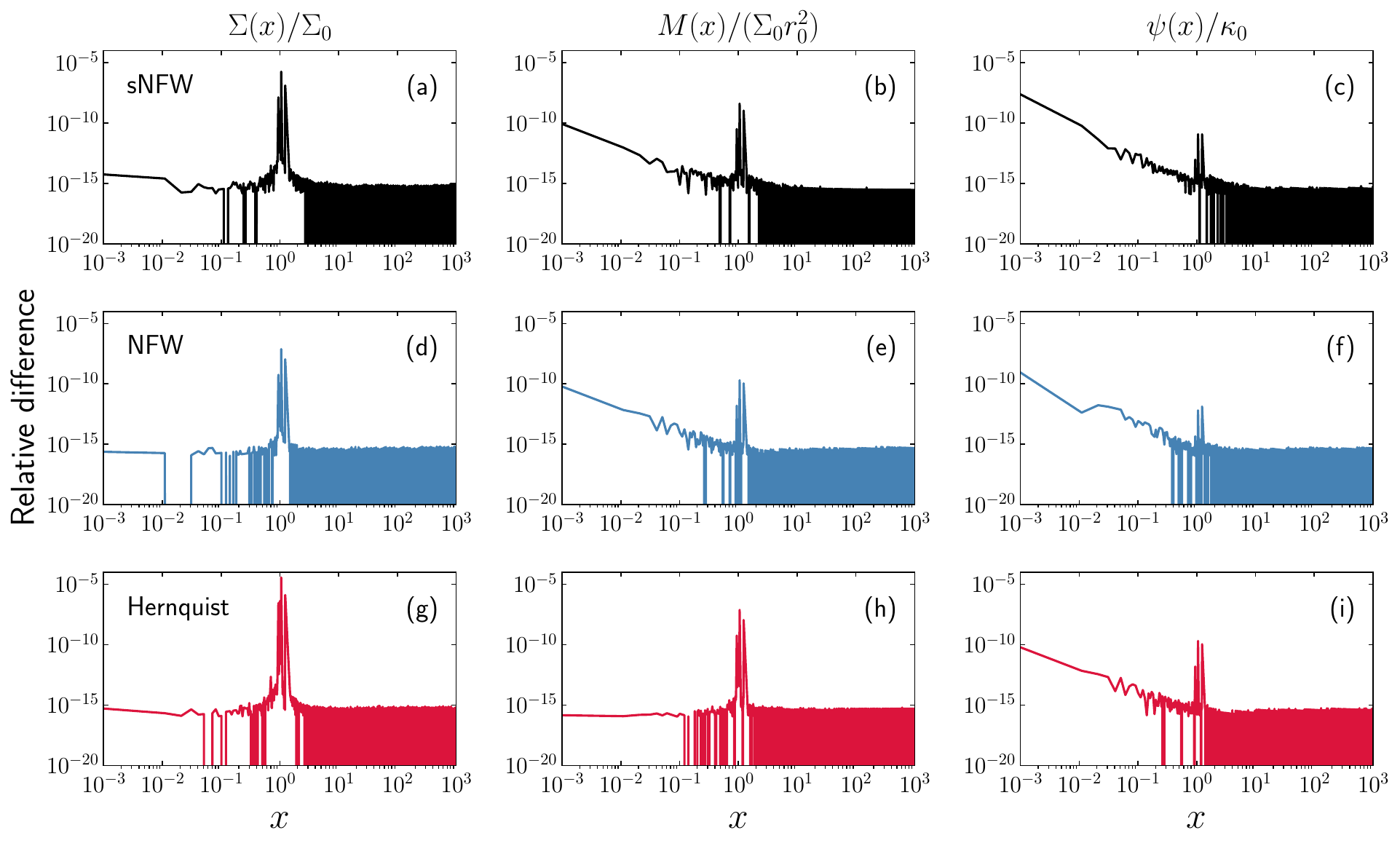}
\caption{Relative difference $|(\text{series}-\text{analytical})/\text{analytical}|$ given for the normalised surface mass density (left-hand column), normalised projected mass (middle column), and normalised deflection potential (right-hand column). We show the relative difference for the sNFW ($n=7/2$, black, upper row), NFW ($n=3$, blue, middle row), and Hernquist ($n=4$, red, lower row) lenses.}
\label{fig: relative difference}
\end{figure*}

\subsection{Performance}\label{sec: performance}
As described above, we have simplified and expressed the Fox $H$-functions of interest in terms of simpler functions. Unfortunately, in general this is not possible, leaving the corresponding power series representations as a practical alternative. With this in mind, in this section we assess the accuracy and performance of the power series given in Sec. \ref{sec: series representation}, demonstrating the value of working with the Fox $H$-function.

We start writing each power series in the form
\begin{equation}\label{eq: SN}
S_N(x)\approx\sum_{k=0}^{N}C_{k}(x),
\quad\text{with}\quad
S_{N+1}(x)=S_{N}(x) + C_{N+1}(x),
\end{equation}
where $C_{k}(x)$ denotes the $k$-th term of the given power series, and $S_{N}(x)$  represents the summation of the first $N+1$ terms. For instance, for the power series \eqref{eq: series smd 0<x<1} we can write
\begin{equation}\label{eq: ck}
C_k(x)=\dfrac{2^{n-3}}{\pi|x|}a_{1,k}x^{2k}\left(a_{2,k}-\ln\left(x^2\right)\right),
\end{equation}
where we use the identities $\Gamma(k+1)=k\Gamma(k)$ and $\psi_{0}(k+1)=\psi_0(k)+1/k$ to  express the coefficients $a_{1,k}$ and $a_{2,k}$ recursively (for convenience) as
\begin{equation}\label{eq: ak1}
a_{1,k} = 
\begin{dcases}
\Gamma\left(\dfrac{n-1}{2}\right)\Gamma\left(\dfrac{n}{2}\right)\quad & \text{if } k=0 \\[3 mm]
\dfrac{(2k + n - 3)(2k + n - 2)}{4k^2}a_{1,k-1}\quad & \text{if } k>0
\end{dcases}
\end{equation}
and
\begin{equation}\label{eq: ak2}
a_{2,k} = 
\begin{dcases}
2\psi_{0}(1)-\psi_0\left(\dfrac{n-1}{2}\right) - \psi_0\left(\dfrac{n}{2}\right)\quad & \text{if } k=0 \\[3 mm]
\dfrac{2}{k} - \dfrac{2}{2k + n - 3} - \dfrac{2}{2k + n - 2} + a_{2,k-1} \quad & \text{if } k>0
\end{dcases}.
\end{equation}

Now, the question is: when should we stop the summation? In this regard, we adopt an approach similar to the one used for the  Gauss hypergeometric function in \cite{pearson2017numerical}. We stop the summations once the three conditions $|C_N(x)/S_{N-1}(x)|< \text{tol}$, $|C_{N-1}(x)/S_{N-2}(x)|< \text{tol}$, and $|C_{N-2}(x)/S_{N-3}(x)|< \text{tol}$ are satisfied simultaneously for a given tolerance (tol).

Considering $n=3$ (NFW), Fig. \ref{fig: N vs x} depicts the number of terms required for each summation to converge ($N_{\text{max}}+1$) as a function of $x$ for the three Fox $H$-functions of interest, given a tolerance of $10^{-15}$ (which we use in this work unless otherwise stated). Since such functions are even, we restrict the domain in Fig. \ref{fig: N vs x} to $x>0$. All three functions exhibit a similar behaviour, where the summations require about $10$ terms when $x\ll 1$ and $x\gg 1$, while this number increases rapidly as $x \to 1$. This is expected, considering the discontinuity at $x^2 = 1$ in all three functions. For $n=7/2$ (sNFW) and $n=4$ (Hernquist) the number of terms required in each summation is also consistent with Fig. \ref{fig: N vs x}.

The imposed stopping conditions imply that $N_{\text{max}}$ is unknown. Thus, every time we evaluate a Fox $H$-function, we must verify whether these stopping conditions are met, and we must repeatedly compute coefficients such as $a_{1,k}$ and $a_{2,k}$ in \eqref{eq: ck},  even though they are independent of $x$. Therefore, this way of computing the different summations involved is accurate, but time consuming. 

We can improve the computation time simply by fixing $N_{\text{max}}$, which we can infer from the information given in Fig. \ref{fig: N vs x}. Nonetheless,  in our case, we need to be careful because of how $N_{\text{max}}$ behaves close to $x^2 = 1$. Then, if the domain of interest is not close to $x^2 = 1$, it is possible to choose a global $N_{\text{max}}$. Otherwise, we can divide the domain into disjoint subsets and fix a different $N_{\text{max}}$ for each subset. Therefore, from Fig. \ref{fig: N vs x} and with $n=3$ (NFW), $n=7/2$ (sNFW), and $n=4$ (Hernquist) in mind, for the three Fox $H$-functions of interest we take $N_{\text{max}}=50$ when $0<x^2<0.5$ or $x^2\geq 1.5$, $N_{\text{max}}=200$ when $0.5\leq x^2<0.9$ or $1.1\leq x^2<1.5$, and $N_{\text{max}}=10^4$ when $0.9\leq x^2<1$ or $1<x^2<1.1$. Thus, with $N_{\text{max}}$ in hand,  we need to compute $a_{1,k}$ and $a_{2,k}$ only once, which leads to faster summations. 

Now, in order to assess the performance of our approach, we use it to compute the lensing properties such as the surface mass density, projected mass, and deflection potential for the NFW, sNFW, and Hernquist lenses as they are given in equations  \eqref{eq: gNFW surface mass density H}-\eqref{eq: gNFW deflection potential H}. Then, we compare their computation times ($t_{\text{series}}$) against the results ($t_{\text{numerical}}$) obtained through direct numerical integration (where we use the method quad from \textsc{scipy}) applied to the lensing properties as defined in equations \eqref{eq: surface mass density}, \eqref{eq: projected mass}, and \eqref{eq: deflection potential}. For both \eqref{eq: projected mass} and \eqref{eq: deflection potential}, we  use the analytical representation of $\Sigma(x)$ obtained from \eqref{eq: gNFW surface mass density H} and \eqref{eq: smd h hypergeometric} (as appropriate) instead of using the result of the numerical integration in \eqref{eq: surface mass density}. We do this because, for the surface mass density,  it is usually possible to find an analytical expression. The ratio $t_{\text{numerical}}/t_{\text{series}}$ is depicted in Fig. \ref{fig: computation time performance} (upper row) as a function of the dimension less coordinate $x$, where we can see that, in general, for the surface mass density (panel a) $t_{\text{numerical}}/t_{\text{series}}\gtrsim 10$, while for the projected mass (panel b) and deflection potential (panel c), $t_{\text{numerical}}/t_{\text{series}}\gtrsim 50$, where all three lenses show a similar behaviour. Overall, the power series approach outperforms the direct numerical integration; however, the power series performance drops when we get close to $x=1$ (vertical dashed lines). This behaviour is expected, considering the increase in the number terms needed for the power series to achieve a given tolerance.

Now, Fig. \ref{fig: computation time performance} (lower row) depicts the ratio $t_{\text{numerical}}/t_{\text{analytical}}$, which compares the computation time of the direct numerical integration with the analytical counterpart ($t_{\text{analytical}}$), which correspond to the expressions listed in Sec. \ref{sec: final expressions}. We use the  implementation of Gauss hypergeometric function in \textsc{scipy}. In all three panels (d - f), we can see that, even though the analytical expressions are not given (in general) in terms of simple functions, they also outperform the numerical integration. However, for the projected mass (panel e) and deflection potential (panel f), the NFW ($n=3$, blue) shows a better performance compared to the other two lenses. This behaviour is a consequence of $H^{2,2}_{3,3\,M}\left(x^2,n=3\right)$ \eqref{eq: h pm final expression n = 3} and $H^{2,3}_{4,4\,\psi}\left(x^2,n=3\right)$ \eqref{eq: h dp final expression n = 3} being given in terms of inverse trigonometric and hyperbolic functions instead of the Gauss hypergeometric function (as justified in Sec. \ref{sec: final expressions}). With that being said, when no particular index $n$ is of interest, the general results in Sec. \ref{sec: final expressions} are practical. 

If performance is a priority and it is clear what model to use (which index $n$), the best option is to take an extra step, and simplify (as much as possible) the required Fox $H$-functions using the results of either Sec. \ref{sec: series representation} or Sec. \ref{sec: final expressions}, aiming for expressions given in terms of more standard functions, which are usually easier to manipulate and faster to compute. For the NFW and Hernquist lenses such expressions are already known and are listed in Appendix \ref{appendix: nfw and hernquist}, whereas for the sNFW we provide the simplified expressions in Sec. \ref{sec: sNFW lensing}.

Accuracy is also a relevant aspect when performing calculations. Here, we evaluate the accuracy of computing the lensing properties in terms of the power series representation by comparing with the corresponding simplified analytical expressions, where we use the relative difference $|(\text{series}-\text{analytical})/\text{analytical}|$. In Fig. \ref{fig: relative difference} we show the relative difference for the normalised surface mass density (left-hand column), normalised projected mass (middle column), and normalised deflection potential (right-hand column). For all three quantities, and for the three lens models as well, the highest relative difference (at most about $10^{-5}$) occurs close to either $x=0$ or $x=1$, where as we know, our Fox $H$-functions are not well defined. Nonetheless, in general the relative difference is small, with a value of around $10^{-15}$  when we move away from $x=0$ and $x=1$. Therefore, there is no doubt that working with the power series representations of the Fox $H$-function is a good option, considering that when properly manipulated (as with any other power series), they provide accurate and considerably fast calculations.


\section{The super-NFW as a gravitational lens}\label{sec: sNFW lensing}
\begin{figure*}
\centering
\includegraphics[width=0.9\linewidth]{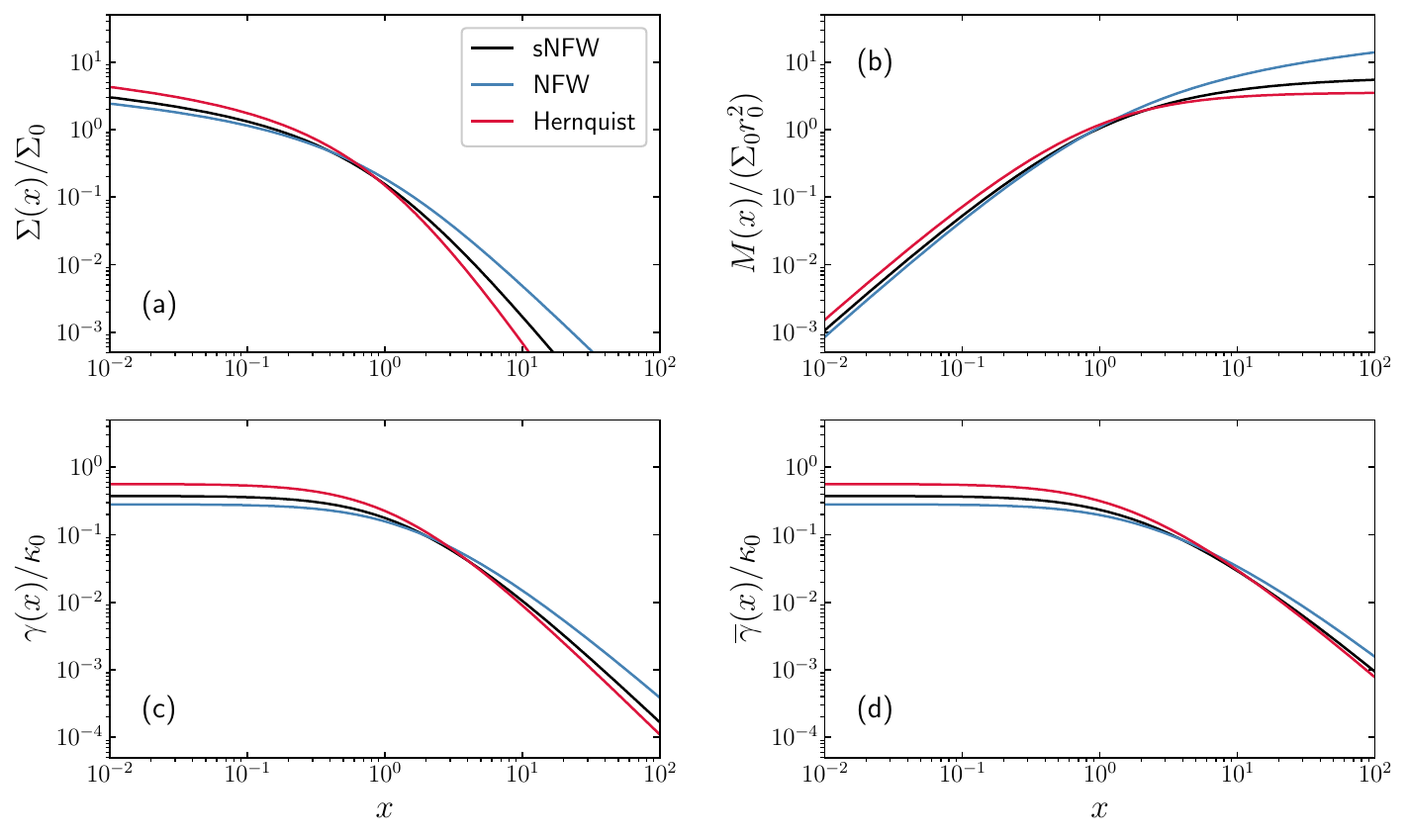}
\caption{
(a) Normalised surface mass density, (b) normalised projected mass, (c) normalised shear, and (d) normalised  mean shear. In all panels the curves are given for the sNFW (black), NFW (blue), and Hernquist (red) lenses.
}
\label{fig: all}
\end{figure*}
\begin{figure*}
     \centering
     \includegraphics[width=0.9\linewidth]{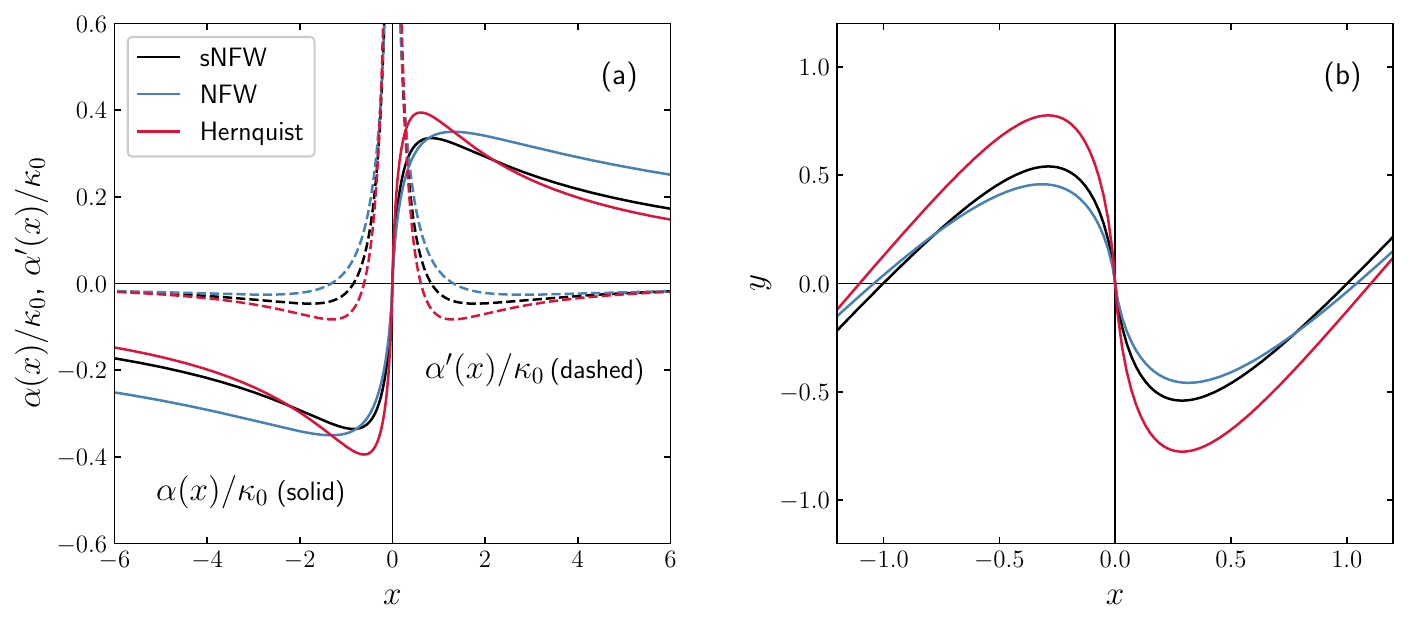}
     \caption{
     (a) Normalised deflection angle (solid curve) and its corresponding derivative (dashed curve). (b) Lens equation \eqref{eq: deimensionless lens equation 1d} given for $\kappa_0=3$. In both panels the curves are given for the sNFW (black), NFW (blue), and Hernquist (red) lenses.
     }
     \label{fig: deflection angle}
\end{figure*}

Considering the gNFW defined by \eqref{eq: gnfw density}, in Sec. \ref{sec: gNFW lensing} we found analytical expressions for its lensing properties \eqref{eq: gNFW surface mass density H}-\eqref{eq: gNFW deflection potential H}, which are given in terms of the Fox $H$-function. Our goal in this section is to simplify such expressions (using either the power series in Sec. \ref{sec: series representation} or the analytical results in Sec. \ref{sec: final expressions}) for the particular case $n=7/2$, or sNFW as it was named by \cite{lilley2018snfw}. As we argued is Sec. \ref{sec: performance}, this process is advantageous since analytical expressions given in terms of the simplest possible functions are generally faster to compute, and easier to work with.

We start with the surface mass density, which can be expresses as
\begin{equation}\label{eq: snfw surface mass density}
\Sigma(x) = \Sigma_0
\begin{dcases}
\dfrac{9\pi}{128\sqrt{2}} & \text{if } x= 1\\
\dfrac{(3+x)K\left(\frac{1-x}{1+x}\right)-4E\left(\frac{1-x}{1+x}\right)}{2(1-x)^2(1+x)^{3/2}} & \text{if } x\neq 1
\end{dcases},
\end{equation}
where $K$ and $E$ are the complete elliptic integrals of first and second kind, respectively. This result is equivalent to the expression reported by \cite{lilley2018snfw}. For the sNFW it is possible to find its surface mass density directly from \eqref{eq: surface mass density} without the need to use any fancy techniques (as it is the case with many profiles).   If we use the surface mass density as given in \eqref{eq: snfw surface mass density}, and try to find the projected mass and deflection potential from definitions \eqref{eq: projected mass} and \eqref{eq: deflection potential}, the integrals involved become too difficult to be solved by traditional means, or at least it is not evident how to proceed with the integration. Here, is where the use of the Fox $H$ function comes handy. With that being said, for the projected mass and deflection potential the Fox $H$-function leads to 

\begin{equation}\label{eq: snfw projected mass}
M(x) = 2\pi r_0^2 \Sigma_0
\begin{dcases} 1-\dfrac{3\pi}{8\sqrt{2}} & \text{if } x= 1\\
\dfrac{E\left(\frac{1-x}{1+x}\right)-xK\left(\frac{1-x}{1+x}\right)}{\sqrt{1+x}(x-1)} + 1 & \text{if } x\neq 1
\end{dcases}
\end{equation}
and
\begin{equation}\label{eq: snfw deflection potential}
\psi(x) = 2\kappa_{0} \left(\frac{2K\left(\frac{1-x}{1+x}\right)}{\sqrt{1+x}}+ \ln\left(\frac{x}{8}\right)\right),
\end{equation}
respectively. Since we are dealing with $n>3$, according to \eqref{eq: gnfw total mass} the sNFW possesses a finite mass. Together with  \eqref{eq: gnfw surface mass density 0}, the characteristic surface mass density for the sNFW becomes $\Sigma_0=M_{\infty}/(2\pi r_0^2)$, where $M_{\infty}$ represents its total mass. Recall that $\kappa_0 = \Sigma_0/\Sigma_{\text{cr}}$ is the characteristic convergence.

We substitute these results into definitions \eqref{eq: shear} and \eqref{eq: mean shear - potential} (as appropriate), and find the corresponding shear and mean shear, which can be written as
\begin{align}\label{eq: snfw shear}
\nonumber
&\gamma(x) = \kappa_0\\
&\times
\begin{dcases} 
2-\dfrac{105\pi}{128\sqrt{2}}\hspace*{\fill}\qquad\text{if } x= 1\\
\dfrac{x(4-3x-5x^2)K\left(\frac{1-x}{1+x}\right)-4(1-2x^2)E\left(\frac{1-x}{1+x}\right)}{x^2\sqrt{1+x}(x-1)} 
+ \dfrac{2}{x^2} & \\
\hspace*{\fill}\qquad\text{if } x\neq 1
\end{dcases}
\end{align}
and
\begin{align}\label{eq: snfw mean shear}
\nonumber
&\overline{\gamma}(x) = 2\kappa_0\\
&\times
\begin{dcases} \dfrac{19\pi}{8\sqrt{2}} - 6\ln(2) - 1& \text{if } x= 1\\
\dfrac{(5x-4)K\left(\frac{1-x}{1+x}\right)-E\left(\frac{1-x}{1+x}\right)}{x^2\sqrt{1+x}(x-1)} 
+ \dfrac{2}{x^2}\ln\left(\dfrac{x}{8}\right) - \dfrac{1}{x^2} & \text{if } x\neq 1
\end{dcases},
\end{align}
respectively. From \eqref{eq: shear x=0}, we can see that both quantities satisfy $\gamma(0)/\kappa_0 = \overline{\gamma}(0)/\kappa_0=3/8$. Recall that for \eqref{eq: snfw surface mass density}-\eqref{eq: snfw mean shear} we consider $x>0$. Lastly, by substituting \eqref{eq: snfw projected mass} into the general definition \eqref{eq: deflection angle}, the deflection angle takes the form
\begin{equation}\label{eq: snfw deflection angle}
\alpha(x) = \dfrac{2 \kappa_0}{x}
\begin{dcases} 1-\dfrac{3\pi}{8\sqrt{2}} & \text{if } |x|= 1\\
\dfrac{E\left(\frac{1-|x|}{1+|x|}\right)-|x|K\left(\frac{1-|x|}{1+|x|}\right)}{\sqrt{1+|x|}(|x|-1)} + 1 & \text{if } |x|\neq 1
\end{dcases},
\end{equation}
where the discontinuity at $x=0$ is removable, with $\alpha(0)=0$.

From Fig. \ref{fig: all} (panel a) it is evident that the surface mass density for the sNFW falls off asymptotically faster than for the NFW, but slower than for the Hernquist lens, which is consistent with how the profile \eqref{eq: gnfw density} behaves. As a result, unlike the NFW, whose projected mass (Fig. \ref{fig: all} panel b) increases logarithmically and eventually diverges towards the outskirts, for the sNFW and Hernquist lenses, their projected mass tends asymptotically to a finite value (total mass), with the smallest being that of the Hernquist lens. Now, the distortions induced by the lens are stronger within the regions of higher density, thus, for the sNFW the shear and mean shear signals (Fig. \ref{fig: all} panel c and d, respectively) decrease faster than those of the Hernquist lens. Something similar happens for the deflection angle, as it is depicted in Fig. \ref{fig: deflection angle} (panel a). I summary, we have that towards the outskirts, the lensing properties for the sNFW satisfy the condition of being in between properties of the NFW and Hernquist lenses, in agreement with the motivation for defining the sNFW as discussed in \cite{lilley2018snfw}.


\subsection{Image formation and magnification}

\begin{figure*}
     \centering
     \includegraphics[width=0.8\linewidth]{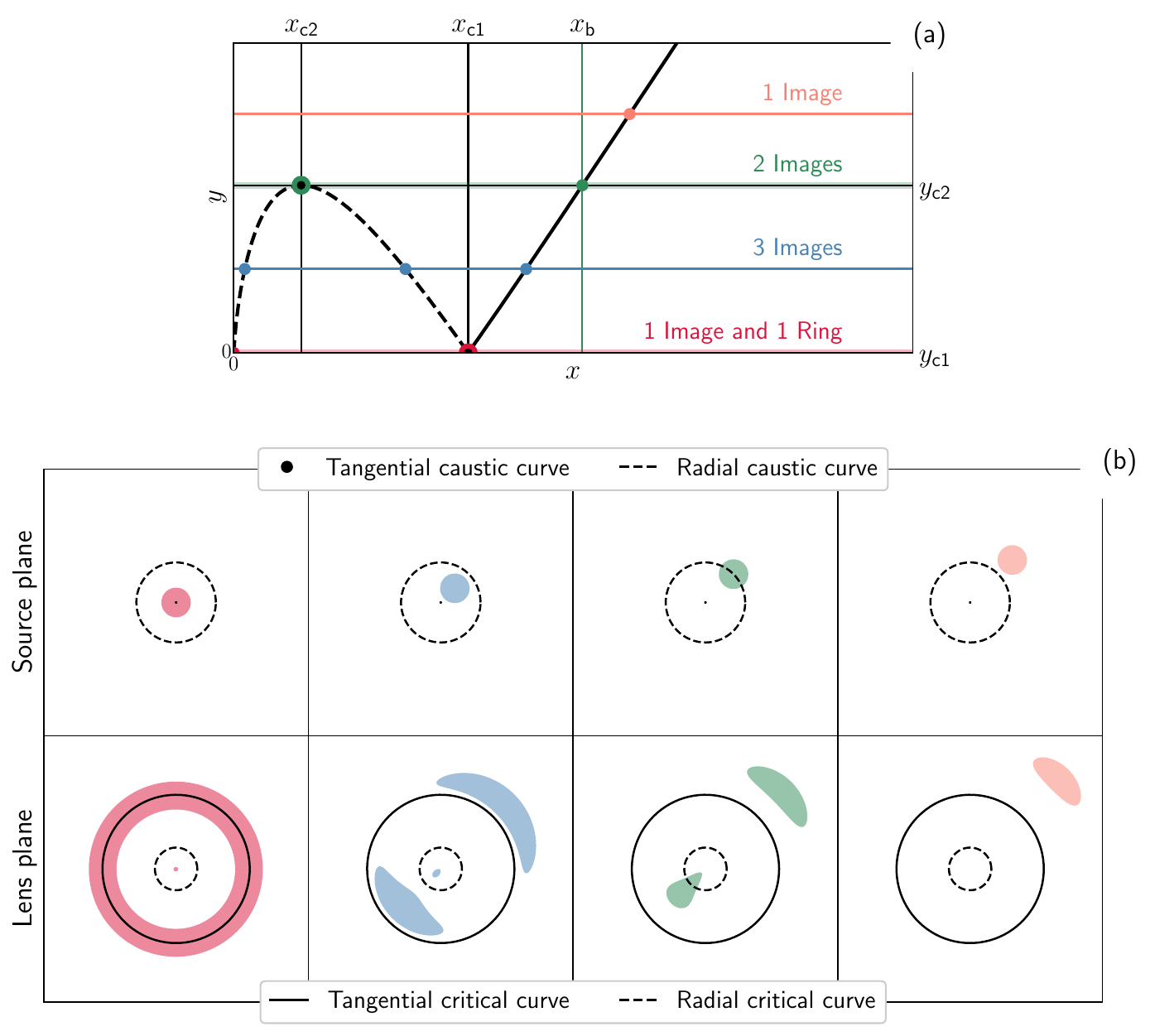}
     \caption{
     Image formation anatomy. In panel (a) the black curve represents the lens equation, such that for a source at position $\boldsymbol{y}=(y\geq0, \varphi)$, it is possible to obtain images at position $\boldsymbol{x}=(x\geq0, \varphi)$ (solid curve) or at position $\boldsymbol{x}=(x\geq0, \varphi+\pi)$ (dashed curve). The vertical and horizontal black lines represent the position of critical and caustic curves, respectively. The additional horizontal lines show the possible scenarios for image formation depending on $y$, while the vertical green line at $x_{\text{b}}$ represents the boundary within which multiple images can appear. Additionally, panel (b) shows an example of image formation for a sNFW lens, and an extended source with different positions relative to the caustic curves. Each colour matches one of the four scenarios described in panel (a).
     }
     \label{fig: image formation anatomy}
\end{figure*}
\begin{figure*}
     \centering
     \includegraphics[width=0.9\linewidth]{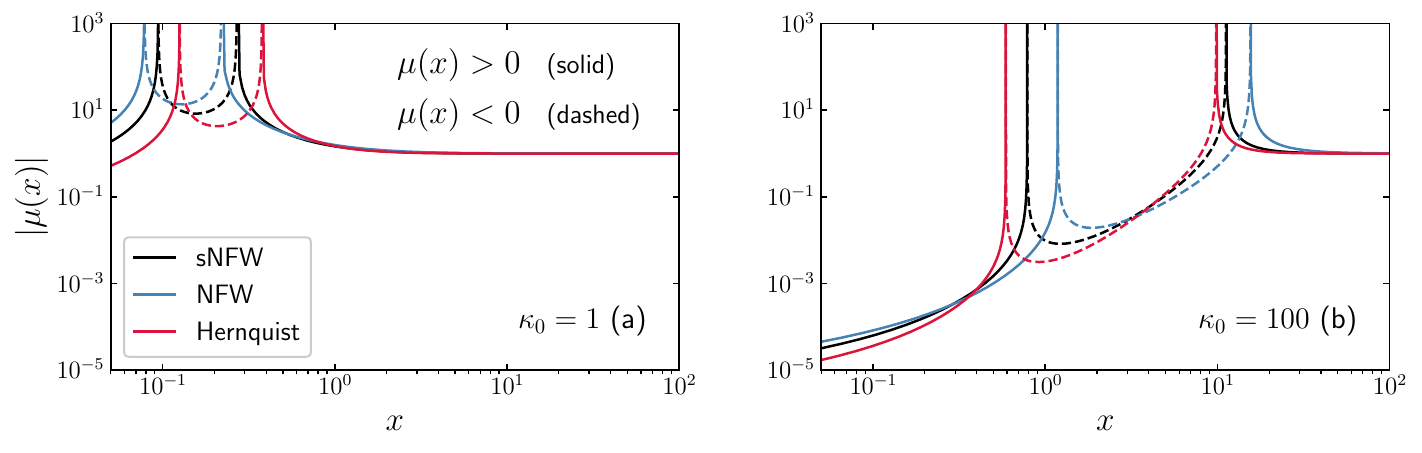}
     \caption{
     Absolute magnification $|\mu(x)|$  given for (a) $\kappa_0=1$ and (b) $\kappa_0=100$. In both panels the curves are given for the sNFW (black), NFW (blue), and Hernquist (red) lenses, where the solid and dashed curves correspond to $\mu(x)>0$ and $\mu(x)<0$, respectively.
     }
     \label{fig: magnification}
\end{figure*}

The non-linear nature of the lens equation opens the possibility of obtaining multiple images of the same source. Here, for the sNFW, we discuss how many images can be formed, and under which conditions. In that regard, Fig. \ref{fig: deflection angle} (panel b) depicts the lens equation with $\kappa_0=3$ for the sNFW (black), NFW (blue), and Hernquist (red) lenses, where it is clear that for these three models the features and behaviour of the lens equation are overall the same. For different values of $\kappa_0$ the main changes occur at the intersections (there are two of them) between the lens equation and the x-axis, which move closer or farther from the origin as $\kappa_0$ decreases or increases, respectively. Also, variations in $\kappa_0$ induce changes in amplitude, such that when the inner concave and convex sections of the lens equation disappear, it is no longer possible to obtain multiple images.   
Note that for many axisymmetric lenses their lens equation behaves similarly, with the exception of, for example, the point-like and SIS lenses (see e.g., \cite{schneider2006gl}). 
 
All information regarding image formation is summarised in Fig. \ref{fig: image formation anatomy} (panel a), where for simplicity, we restrict the radial coordinates to non-negative values. Consider a point-like source at position $\boldsymbol{y}=(y,\varphi)$ on the source plane, where $\varphi$ is its azimuthal coordinate. The black curve represents the lens equation, where the solid part corresponds to possible images on the lens plane at position $\boldsymbol{x}=(x,\varphi)$, while the dashed part corresponds to possible images at position $\boldsymbol{x}=(x,\varphi+\pi)$. The latter case is equivalent to the scenario in Fig. \ref{fig: deflection angle} (panel b) for $x<0$ and $y\geq 0$. 

The critical curves, at which the magnification $\mu(x)$ diverges, rule the presence of multiple images. The magnification is defined as
\begin{equation}
\dfrac{1}{\mu(x)} = (1-\kappa(x))^2 - \gamma^2(x)=\dfrac{y(x)y^{\prime}(x)}{x},
\end{equation}
which leads to two possible critical curves. On the one hand, we have the tangential critical curve. This curve is a circle (also known as the Einstein ring) with radius $x_{\text{c1}}$, which satisfies
\begin{equation}\label{eq: tangential critical curve eq}
y(x)/x=1-\overline{\kappa}(x)=0, 
\quad\text{with}\quad
\overline{\kappa}(x)=\dfrac{\overline{\Sigma}(x)}{\Sigma_{\text{cr}}}= \dfrac{\alpha(x)}{x}.
\end{equation}
This is basically the lens equation applied to a source with radial coordinate $y=0$, but with the restriction $x\neq 0$. Although $x=0$ is a solution to the lens equation, it does not represent a critical curve. On the other hand, we have the radial critical curve. This curve is a circle with radius $x_{\text{c2}}$, which satisfies 
\begin{equation}\label{eq: radial critical curve eq}
y^{\prime}(x)=1-\alpha^{\prime}(x)=0.
\end{equation}
Here, by applying the Leibniz integral rule to the integral representation of $\alpha(x)$ in \eqref{eq: deflection angle}, we obtain $\alpha^{\prime}(x)=2\kappa(x)-\overline{\kappa}(x)$. Thus, $x=0$ cannot represent a critical radial curve either.

Once we map the critical curves into the source plane, we get the caustic curves. These curves delimit the regions where a source will lead to multiple images. The tangential critical curve is mapped into $y_{\text{c1}}=0$, while the radial critical curve is mapped into a circle with radius $y_{\text{c2}}$. Critical and caustic curves are concentric. In Fig. \ref{fig: image formation anatomy} (panel a) the position of the critical curves and caustic curves are represented as vertical and horizontal solid black lines, respectively. For illustration, in Fig. \ref{fig: image formation anatomy} (panel b) we show an example of these curves for a sNFW lens with $\kappa_0=3$.

Now, since $x_{\text{c1}}>x_{\text{c2}}$, the strong lensing effect can be produced as long as the tangential critical curve is present, which occurs when the characteristic convergence $\kappa_0$ is greater than a certain minimum ($\kappa_{0, \text{min}}$). Given that the tangential critical curve satisfies $\overline{\kappa}(x) = \alpha(x)/x= 1$, once we make explicit $\alpha(x)$ and write $\kappa_0$ as a function of $x$, $\kappa_{0, \text{min}}$ is the result of evaluating the limit when $x\to 0$ (no tangential critical curve is produced). In our case, from \eqref{eq: deflection angle} and \eqref{eq: gNFW projected mass H}, for the gNFW lens we get
\begin{equation}
\kappa_0=1/\left(x H^{2,2}_{3,3\,M}\left(x^2,n\right)\right). 
\end{equation}
Thus, from either the power series representation \eqref{eq: series pm 0<x<1}, or the representations \eqref{eq: h pm final expression n neq 3} (for $n\neq 3$) and \eqref{eq: h pm final expression n = 3} (for $n=3$), it can be shown that $\kappa_{0, \text{min}}=0$. Therefore, any lens described by the gNFW produces a tangential critical curve if $\kappa_0>0$. 

This result, however, does not apply to all profiles. For instance, if we consider a Non-Singular Isothermal Sphere (NIS) with core radius $r_c\neq 0$, from the expressions listed in Appendix \ref{appendix: sis and nis} we can show that in this case $\kappa_{0, \text{min}}=1/\sqrt{\pi}$. Note that $\kappa_{0, \text{min}}$ is subject to the definition of $\kappa_0$, in particular when $\kappa_{0, \text{min}}\neq0$. 

Returning to image formation, from Fig. \ref{fig: image formation anatomy} (panel a) we can see that for a source at $y=0$ (horizontal red line) there are two possible solutions to the lens equation. These solutions correspond to the critical curve $x_{\text{c1}}$ and one image at the centre ($x=0$). At $x=0$ the magnification is finite and, for lenses with a divergent surface mass density like the gNFW, we have $\mu(x \to 0) = 0$. Such demagnified/faint image is difficult (if not impossible) to observe. For sources located at $0<y<y_{\text{c2}}$ (horizontal blue line) we get three images. One of them is located at $x>x_{\text{c1}}$, while the other the two are located at $0<x<c_{\text{c1}}$. Those two images merge at $x_{\text{c2}}$ when the given source is located at $y_{\text{c2}}$, for which a second image is formed at $x>x_{\text{c1}}$ (horizontal green line). For sources located at $y>y_{\text{c2}}$ (horizontal orange line) only one image is formed at $x>x_{\text{c1}}$. Recall that, if the source has an azimuthal coordinate $\varphi$, the images with radial coordinate $0<x<c_{\text{c1}}$ have azimuthal coordinate $\varphi + \pi$, while the images with radial coordinate $x>x_{\text{c1}}$ have azimuthal coordinate $\varphi$. As a final remark, in Fig. \ref{fig: image formation anatomy} (panel b) we illustrate these four possible scenarios of image formation for an extended source with radius $0.2$ and a sNFW lens with $\kappa_0=3$.

Fig. \ref{fig: magnification} shows the absolute magnification $|\mu(x)|$ for the sNFW (black), NFW (blue), and Hernquist (red) lenses. These curves are given for  $\kappa_0=1$ (panel a) and $\kappa_0=100$ (panel b). In terms of signed magnification, the solid and dashed curves represent $\mu(x)>0$ and $\mu(x)<0$, respectively. The morphological features exhibited by these curves are essentially the same for the three lenses of interest. The most evident similarity corresponds to two prominent vertical asymptotes, which represent the position of the critical curves. The inner one represents the radial critical curve, while the outer one represents the tangential critical curve. Therefore, images located at $x_{\text{c2}}<x<x_{\text{c1}}$ have negative parity, while images located at $0<x<x_{\text{c2}}$ and $x> x_{\text{c1}}$ have positive parity. Also, we can see that, independently of $\kappa_0$, the magnification satisfies $\mu(x)>1$ for $x>x_{\text{c1}}$ (as expected). When comparing the magnifications, we can see that in general, in terms of magnitude, the magnification produced by the sNFW is larger than that of the Hernquist, but smaller than that of the NFW. The differences in magnitude are particularly evident within $x_{\text{c2}}<x<x_{\text{c1}}$. Moreover, there is a clear dependence on $\kappa_0$. We can see that, within the region $0<x<x_{\text{c1}}$, for these three models $|\mu(x)|$ decreases as $\kappa_0$ increases, producing demagnified images unless they are formed close to a critical curve. 

\begin{figure*}
     \centering
     \includegraphics[width=0.9\linewidth]{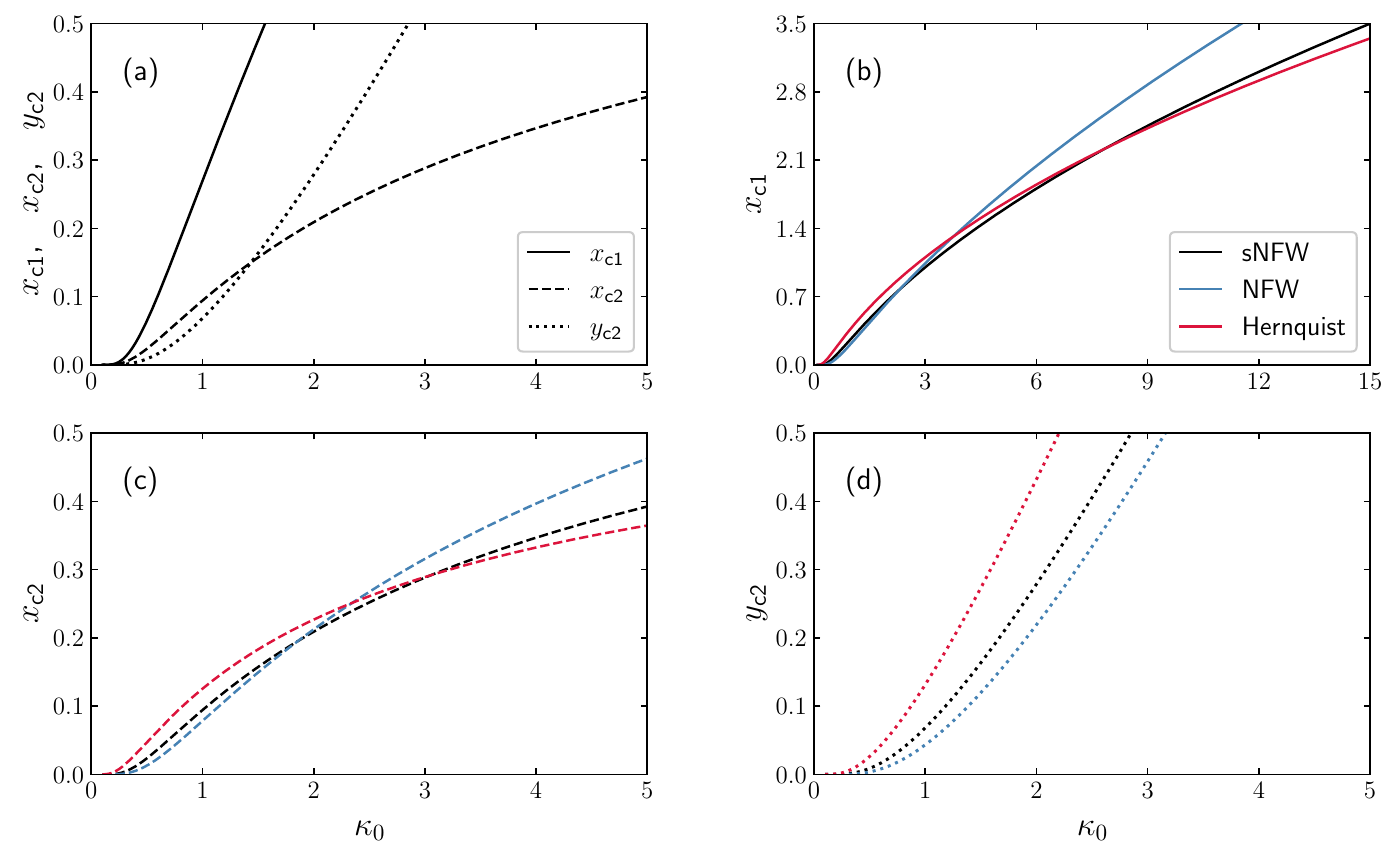}
     \caption{
     For the sNFW, panel (a) shows its tangential critical curve $x_{\text{c1}}$ (solid), radial critical curve $x_{\text{c2}}$ (dashed), and radial caustic curve $y_{\text{c2}}$ (dotted) as a function of the characteristic convergence $\kappa_0$. Additionally, panels (b)-(d) depict each of these curves for the sNFW (black), NFW (blue), and Hernquist (red) lenses.  Note that for these lenses the tangential critical curve is formed as long as $\kappa_0>0$, and its corresponding caustic is $y_{\text{c1}} = 0$.
     }
     \label{fig: critical and caustic curves}
\end{figure*}

Another relevant aspect to take into account, is how the critical and caustic curves behave as a function of $\kappa_0$. In this regard,  Fig. \ref{fig: critical and caustic curves} (panel a) shows these curves simultaneously for the sNFW lens. We can see that the critical curves move away from the centre, and the separation between critical curves increases, always satisfying $x_{\text{c1}}>x_{\text{c2}}$. We can also see that $x_{\text{c2}}>y_{\text{c2}}$ for small values of $\kappa_0$, however, eventually this relation is reversed and we get $y_{\text{c2}}>x_{\text{c2}}$. For higher values of $\kappa_0$, the radial critical and caustic curves keep increasing in size, but the radial caustic grows faster. This implies that when $\kappa_0$ increases it becomes easier to get multiple images, and since in terms of proportion the region within the radial critical curve is smaller, it is more likely to get a faint image close to the centre. 

For the sNFW, NFW, and Hernquist lenses, the curves $x_{\text{c1}}$, $x_{\text{c2}}$, and $y_{\text{c2}}$ as a function of $\kappa_0$ have a similar behaviour. In this regard, in Fig. \ref{fig: critical and caustic curves} (panels b and c) we can see that for small values of $\kappa_0$, the Hernquist lens produces larger critical curves than the other two lenses, but as $\kappa_0$ increases, the NFW takes its place and its critical curves start growing faster than those of the other two lenses (with the sNFW being in between). In particular, up to $\kappa_0\sim 2$ the critical curves are comparable in size for the sNFW and NFW lenses, with the sNFW producing slightly larger curves, but as $\kappa_0$ increases they start getting apart from each other and this time the NFW produces significantly larger curves.

With respect to the caustic curves, for the three models the tangential caustic curve is always $y_{\text{c1}}=0$, independently of $\kappa_0$. For the radial caustic curves, we can see in Fig. \ref{fig: critical and caustic curves} (panel d) that, independently of $\kappa_0$, for the Hernquist lens its radial caustic curve is always larger and grows faster than those of the other two lenses, while the NFW  always produces the smallest radial caustic curves. These results imply that it is much easier for the Hernquist or sNFW lenses to produce the strong lensing effect than it is for the NFW lens.

As a final remark on image formation, provided that sources are not located on caustics ($y\neq 0$ and $y\neq y_{\text{c2}}$), we get either one or three images, in agreement with the odd images theorem. First explored in spherical galaxies by \cite{dyer1980ApJmultipleimages}, and then generalised by \cite{burke1981theoren}, it requires lenses to have a non-singular mass  distribution. However, it was shown by \cite{bartelmann1996NFWarcs} that a mass distribution such as the NFW, despite being singular at its centre, possesses a radial critical curve, forming an odd number of images. He argues that this is possible since $\alpha^{\prime}(x)$ is continuous (for $x > 0$), $\alpha^{\prime}(x\to 0)=\infty$, and $\alpha^{\prime}(x\to \infty)=0$. This guaranties that there exists an $x$ that satisfies \eqref{eq: radial critical curve eq}. From the dashed curves in Fig. \ref{fig: deflection angle} (panel a) it is straightforward to see that these conditions are also fulfilled by the sNFW and Hernquist lenses. Additionally, as discussed in \cite{mollerach2002gravitational}, the appearance of an odd number of images can be argued from considering the deflection angle being bounded and continuous at $x=0$. Alternatively, from the perspective of the deflection potential, theorem 1 in \cite{petters2010multipleimages} indicates that the total number of images is given by $N=2N_{+}+g-1$, where $N_{+}$ is the number of images with positive parity, and $g$ is the number of infinite singularities of $\psi$, which are those $x_s$ for which $\psi(x\to x_s)=-\infty$. In our case, since $g=0$, this theorem predicts $N=3$ for sources with $0<y<y_{\text{c2}}$, and $N=1$ for sources with $y>y_{\text{c2}}$. Consequently, this suggests that the condition of the mass distribution being non-singular is more a sufficient than a necessary condition for obtaining an odd number of images.

\subsection{Magnification invariant}
\begin{figure*}
     \centering
     \includegraphics[width=0.9\linewidth]{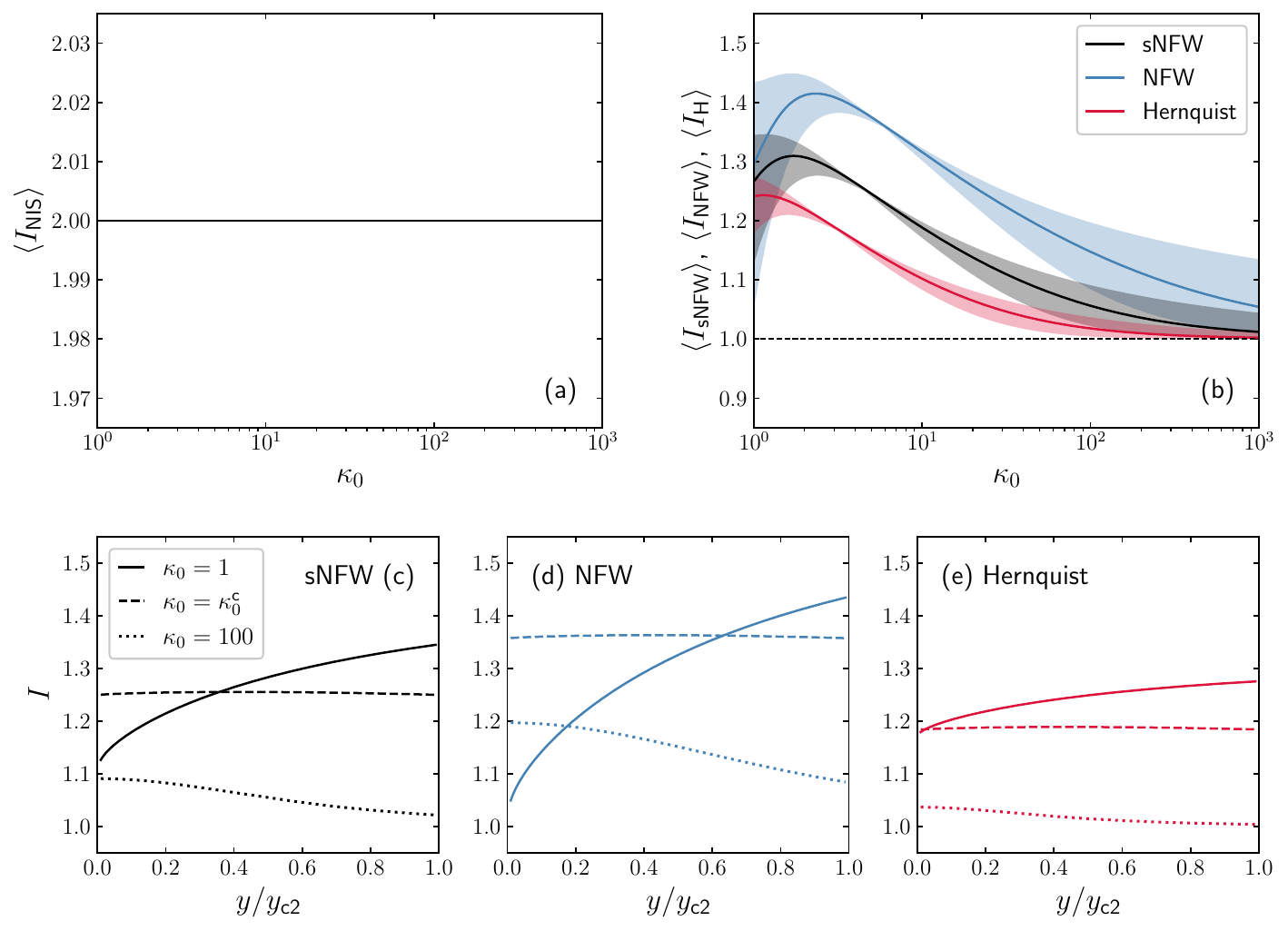}
     \caption{
     For a given characteristic convergence $\kappa_0$, we evaluate $I$ for multiple sources with radial position $0<y<y_{\text{c2}}$ and compute the corresponding average $\braket{I}$. We repeat this process for several values of $\kappa_0$. In panels (a) and (b) the solid curves denote $\braket{I}$, while the corresponding coloured bands enveloping these curves represent the range of possible values that $I$ can have. On the one hand, in panel (a) we show $\braket{I}$ as a function of $\kappa_0$ for the NIS, where it is clear that $\braket{I_{\text{NIS}}}=2$ and the variations in $I$ are negligible (no coloured band), in agreement with the analytical result. On the other hand, in panel (b) we have $\braket{I}$ as a function of $\kappa_0$ for the sNFW (black), NFW (blue), and Hernquist (red). Here, the behaviour is more dynamical, which shows that $I$ not only depends on $\kappa_0$. We can see the variations in $I$ for these three lenses in panels (c)-(e), where we show $I$ as function of $y/y_{\text{c2}}$ for $\kappa_0=1$ (solid), $\kappa_0=\kappa_0^{\text{c}}$ (dashed), and $\kappa_0=100$ (dotted). $I$ is constant for $\kappa_0=\kappa_0^{\text{c}}$, however, the small deviations from a constant line that can be observed, are the result of the uncertainties in finding $\kappa_0^{\text{c}}$.
     }
     \label{fig: magnification invariant}
\end{figure*}

When we have a source inside the caustic that produces the maximum possible number of images, $I=\sum_{i}\mu_i$ represents the sum of their corresponding (signed) magnifications. It has been shown that for certain simple lenses, $I$ is constant and independent of most of their parameters. Therefore, it is usually referred to as the magnification invariant. That is the case, for example, of a system of n-point lenses \citep{witt1995magnification_invariant,rhie1997n_point}, and the Plummer lens \citep{werner2006plummer}, for which $I=1$, or the isothermal sphere for which $I=2$ \citep{dalal1998magnification_invariant, wei2018magnification_invariant}. The magnification invariant has also been analytically computed for models intended to fit quadruple lenses, such as variations of the SIS model \citep{dalal1998magnification_invariant, witt2000magnifications_invariant,dalal2001magnification_invariant,hunter2001magnification_invariant}. More recently, \cite{wei2018magnification_invariant} have explored how $I$ behaves for axisymmetric lenses such as the NFW, Einasto (for $n=1/2,1$), an exponential disc, and the modified Hubble model, where they conclude that for such models $I$ only depends on the characteristic convergence. However, as we discuss below, at least for the NFW, our numerical experiments suggest a more dynamical behaviour.

Here, we study how $I$ behaves for the sNFW, NFW, and Hernquist lenses. For these three models we know that at most three images are formed, which occurs for sources within $0<y<y_{\text{c2}}$. Each of these images appears inside one of three possible regions: region 1 ($0<x<x_{\text{c2}}$), region 2 ($x_{\text{c2}}<x<x_{\text{c1}}$), and region 3 ($x_{\text{c1}} <x<x_{\text{b}}$), where $x_{\text{b}}$ (depicted as the vertical green line in Fig. \ref{fig: image formation anatomy} panel a) represents the outer boundary within which multiple images can form. 

To evaluate $I$,  we need to solve the lens equation \eqref{eq: deimensionless lens equation 1d} accurately. To do so, for a source located at $y_0$, we take each of the three regions and construct a one-dimensional grid  with $N$ nodes and map all of them into the source plane by means of \eqref{eq: deimensionless lens equation 1d}. Then, the node that is mapped closest to $y_0$ corresponds to the image within the corresponding region. The accuracy of this solution can be improved by using this value as an initial guess in some root finding method, as it could be Newton-Raphson. This is the approach applied by \cite{wei2018magnification_invariant}. We opt for a refinement process instead, since the relative differences obtained are in general smaller. Nonetheless, from both approaches the behaviour exhibited by $I$ is the same.  We test this approach and our implementation by applying it to the NIS described in Appendix \ref{appendix: sis and nis},  for which \cite{wei2018magnification_invariant} showed that $I_{\text{NIS}}=2$, independently of any parameter. 

What we do here is, given a $\kappa_0$, we solve the lens equation for multiple sources within $0<y<y_{\text{c2}}$, and compute their corresponding $I$. We then compute its average value $\braket{I}$ and keep track of the variations by means of the maximum and minimum values achieved by $I$. We repeat this process for several values of $\kappa_0$. In Fig. \ref{fig: magnification invariant} (panel a) we can see the results for the NIS, where the solid curve represents $\braket{I_{\text{NIS}}}$, and the variations in $I_{\text{NIS}}$ are represented as a coloured band. From this figure it is clear that $\braket{I_{\text{NIS}}} = 2$ for all $\kappa_0$, and the variations on $I_{\text{NIS}}$ are negligible (no coloured band). In fact, the highest variations with respect to the average that we get are around $10^{-10}$. This behaviour is consistent with the analytical result. Therefore, our implementation works correctly. 

Following the same procedure as for the NIS, we can see in Fig. \ref{fig: magnification invariant} (panel b) that for the sNFW (black), NFW (blue), and Hernquist (red) lenses  $I$ is clearly not constant. We have that for these three lenses, $\braket{I}$ reaches a maximum within $1<\kappa_0<10$, and then, it slowly decreases (showing a strong dependence on $\kappa_0$), in agreement with the results provided by \cite{wei2018magnification_invariant} for the NFW. However, our results differ from theirs regarding the constant nature of $I$ for a fixed $\kappa_0$. They concluded that $I$ is approximately constant regardless of the position within the radial caustic. If we focus on the coloured bands in Fig. \ref{fig: magnification invariant} (panel b), in general we can see a dynamical behaviour in $I$, where it is approximately constant only within a small domain. If, for example, we accept variations $|I_{\text{max}} - I_{\text{min}}|$ of at most $1\%$ of the average value, it is fair to say that $I$ is approximately constant when $4.5<\kappa_0<7.8$ for the NFW, $3.5<\kappa_0<6$ for the sNFW, and $2.4<\kappa_0<4.3$ for the Hernquist lens. These intervals can be adjusted to meet observational uncertainties, and with that, it would be possible to use $I$ as an astrophysical tool for model selection, as it is discussed in e.g., \cite{dalal1998magnification_invariant, witt2000magnifications_invariant}. 

For these three lenses, we observe that $I$ is constant only for one $\kappa_0$,  which we denote $\kappa_0^{\text{c}}$. Its value is different for each model, where $\kappa_0^{\text{c}}\approx 5.9$ for the NFW, $\kappa_0^{\text{c}}\approx 4.5$ for the sNFW, and $\kappa_0^{\text{c}}\approx 3.1$ for the Hernquist lens. We can get a grasp on why this occurs from Fig. \ref{fig: magnification invariant} (panels c - e). There, for $\kappa_0=1$ (solid) the minimum value that $I$ can take ($I_{\text{min}}$) occurs when $y\to 0$, whereas its maximum value ($I_{\text{max}}$) occurs when $y\to y_{\text{c2}}$. However, this behaviour reverses when $\kappa_{0}=100$ (dotted), since $I_{\text{min}}$ occurs when $y\to y_{\text{c2}}$ and $I_{\text{max}}$ occurs when $y\to 0$. Therefore, during this transition from a small $\kappa_0$ to a large one, there exists a $\kappa_0=\kappa_0^{\text{c}}$ for which $I_{\text{min}}=I_{\text{max}}$, indicating that $I$ is constant. Furthermore, our numerical experiments suggest that for these three lenses $I_{\text{min}}\to 1$ as $\kappa_0$ increases (dashed horizontal line), and $I\to 1$ as $\kappa_0\to \infty$. 

\section{Summary and conclusions}\label{sec: summary}
In this work, using of the Fox $H$-function we provide analytical expressions for the gravitational lensing properties of the gNFW profile \eqref{eq: gnfw density} proposed by \cite{evansGNFW2006}, where we focus on the sNFW ($n=7/2$) \citep{lilley2018snfw} as an application case. To achieve this, we provide a description of how we can write the projected mass and deflection potential in terms of the Fox $H$-function, whenever we have a lens whose surface mass density belongs to the family of lenses defined in \eqref{eq: general surface mass density}. From these results, the corresponding convergence, deflection angle, shear, and mean shear can be derived straightforwardly. In particular, the lensing properties for the gNFW \eqref{eq: gNFW surface mass density H}-\eqref{eq: gNFW deflection potential H} are given in terms of three independent Fox $H$-functions.

This framework provides a simple way to obtain compact analytical expressions in terms of the Fox $H$-Function for the lensing properties of several axisymmetric lenses. Unfortunately, the Fox $H$-function has not been implemented yet in most programming languages (such as \textsc{python}, which is widely used in astronomy), making it difficult to use. Therefore, an efficient implementation of the Fox $H$-function would be beneficial. However, this task lies beyond the scope of this work. Instead, we compute the Fox $H$-function using its power series representation. A general description of the Fox $H$-function, including the relevant aspects used in this work, can be found in Appendix \ref{appendix: fox}.  

For the gNFW, it was possible to simplify the required Fox $H$-functions in terms of the Gauss hypergeometric function, which is efficiently implemented in, for example, libraries like \textsc{scipy}. This simplification is not always possible (or at least not evident), making the power series representation an advantageous alternative to work with. In this regard, taking the gNFW as an example, in particular for the sNFW, NFW, and Hernquist lenses,  we compared the computation time between the power series approach and the direct numerical integration of the standard definition of the lensing properties. Our results show that when computing the surface mass density, as long as we avoid regions near $x=1$, the power series approach is at least ten times faster than the direct numerical integration. Similarly, for the projected mass and deflection potential the power series approach is at least fifty times faster. Therefore, it is possible for the power series approach to outperforms the standard numerical integration. 

We also tested the accuracy of such series representations, finding a relative difference of at most $10^{-5}$ near $x = 0$ and $x=1$ (where the power series are indeterminate), and a relative difference of about $10^{-15}$ for $x>1$. These results shows that the Fox $H$-function can be used as a reliable alternative to the standard numerical integration.

When a particular index $n$ (model) is of interest, it is convenient to write its lensing properties in terms of simple and standard functions (if possible). These are usually easier to manipulate and faster to compute than either the Fox $H$-function or the Gauss hypergeometric function. We have done this for the sNFW, for which the required Fox $H$-functions reduce to expressions given in terms of the complete elliptic functions of first and second kind $K$ and $E$, respectively. Its surface mass density \eqref{eq: snfw surface mass density} is equivalent to the expression reported by \cite{lilley2018snfw}, while its projected mass \eqref{eq: snfw projected mass} and deflection potential \eqref{eq: snfw deflection potential} (which are fundamental to describe this axisymmetric lens) have not yet been reported in the literature. We also provide explicit analytical expressions for its deflection angle \eqref{eq: snfw deflection angle}, shear \eqref{eq: snfw shear}, and mean shear \eqref{eq: snfw mean shear}. Compared to the NFW and Hernquist lenses, towards the outskirts of the lens, we found that the different lensing properties of the sNFW are between those of the NFW and Hernquist lenses, which is consistent with the expected behaviour of this profile as argued by \cite{lilley2018snfw}.

For sNFW, NFW, and Hernquist lenses, where each produce a maximum of three images, we evaluated the summation of their signed magnification $I$. For these lenses we found that, once we fix $\kappa_0$, in general $I$ exhibits a strong dependence on the position of the source inside the radial caustic. The variations are within a range $I_{\text{min}}\leq I \leq I_{\text{max}}$, where the boundaries depend on $\kappa_0$. If we consider the average value $\braket{I}$, its dependence on $\kappa_0$ shows a behaviour consistent with what \cite{wei2018magnification_invariant} reported for the NFW, but our results differ on how $I$ behaves for a fixed $\kappa_0$. In fact, we found that $I$ is constant only when $\kappa_0=\kappa_0^{\text{c}}$, where $\kappa_0^{\text{c}}\approx 5.9$ for the NFW, $\kappa_0^{\text{c}}\approx 4.5$ for the sNFW, and $\kappa_0^{\text{c}}\approx 3.1$ for the Hernquist lens. However, we can stablish a interval for $\kappa_0$ around $\kappa_0^{\text{c}}$ for which $I$ can be considered approximately constant (constrained by observational uncertainties). For example, considering deviations of at most $1\%$ of the average value, we obtain $4.5<\kappa_0<7.8$ for the NFW, $3.5<\kappa_0<6$ for the sNFW, and $2.4<\kappa_0<4.3$ for the Hernquist lens. Furthermore, our numerical experiments suggest that for these lenses $I_{\text{min}}\to 1$ as $\kappa_0$ increases, and $I\to 1$ as $\kappa_0\to \infty$. These properties are likely to be satisfied by any profile described by the gNFW.

We have shown that, in the context of gravitational lensing, the sNFW represents a reliable alternative to the NFW. In particular, if we consider that their lensing properties present similar behaviours, but with the advantage of the sNFW having a finite total mass. This results make the sNFW a good option for modelling dark matter halos or sub-halos in  astrophysical lenses, not just at galactic scales, but also for galaxy groups and galaxy clusters. Additionally, it would be valuable to extend the analysis by considering an elliptical sNFW or including an external shear. It would also be interesting to study the magnification relations under this setting, and to explore if $I$ maintains its behaviour. However, these ideas lie beyond the scope of this paper, and we leave them for future work.

\section*{Acknowledgements}
L. Casta\~neda was supported by Patrimonio Autónomo - Fondo Nacional de Financiamiento para la Ciencia, la Tecnología y la Innovación Francisco José de Caldas (MINCIENCIAS - COLOMBIA) Grant No. 110685269447 RC-80740-465-2020, projects 69723. We would like the thank the anonymous referee for the thoughtful report, which helped to improve the clarity of the paper.

\section*{Data Availability}
The figures presented in this work can be reproduced with the scripts available at
\href{https://github.com/torres-daniel/Gravitational-lensing-by-the-super-NFW}{https://github.com/torres-daniel/Gravitational-lensing-by-the-super-NFW}.



\bibliographystyle{mnras}
\bibliography{main} 




\appendix
\section{The Mellin transform technique}\label{appendix: mellin}
Given a function $f$, its Mellin transform is an integral transformation defined as
\begin{equation}\label{eq: mellin transform}
F(u) = \mathfrak{M}_{f}(u) = \int_{0}^{\infty} f(x) x^{u-1}dx,   
\end{equation}
where $u\in\mathbb{C}$. Likewise, the inverse Mellin transform of $F$ takes the form
\begin{equation}\label{eq: mellin inverse transform}
f(x) = \mathfrak{M}^{-1}_{F}(x) = \dfrac{1}{2\pi i}\int_{\mathcal{L}} F(u) x^{-u}du,
\end{equation}
with $\mathcal{L}$ being parallel to the imaginary axis in the complex plane.

Now, if $f$ can be written as the Mellin convolution $(f_1\star f_2)(x)$ of two functions, namely $f_1$ and $f_2$, which reads 
\begin{equation}\label{eq: mellin convolution}
f(x)=(f_1\star f_2)(x)=\int_{0}^{\infty} t^{-1}f_1(t)f_2(x/t) dt,
\end{equation}
its Mellin transform turns out to be
\begin{align}
\nonumber
F(u)&=\mathfrak{M}_{f_1\star f_2}(u)\\\nonumber
&= \int_{0}^{\infty} (f_1\star f_2)(x) x^{u-1}dx\\
&= \int_{0}^{\infty}f_1(t)\left(\int_{0}^{\infty}f_2(x/t)x^{u-1}t^{-1}dx\right)dt,
\end{align}
where, if we consider $h=x/t$, it is easy to show that
\begin{equation}\label{eq: mellin transform convolution}
F(u)=\mathfrak{M}_{f_1\star f_2}(u) = \mathfrak{M}_{f_1}(u)\mathfrak{M}_{f_2}(u).
\end{equation}

Finally, by applying the inverse Mellin transform \eqref{eq: mellin inverse transform} to \eqref{eq: mellin transform convolution}, it follows
\begin{equation}\label{eq: mellin transform technique}
f(x) = \mathfrak{M}^{-1}_{F}(x) = \dfrac{1}{2\pi i}\int_{\mathcal{L}}\mathfrak{M}_{f_1}(u)\mathfrak{M}_{f_2}(u)x^{-u}du,
\end{equation}
which smoothly leads to the definition of the Fox $H$-function when $\mathfrak{M}_{f_1}$  and $\mathfrak{M}_{f_2}$ are given as the product of Gamma functions, as we discuss in Appendix \ref{appendix: fox}.  

\section{The Fox $H$-function}\label{appendix: fox}
The Fox $H$-function was first introduced in \cite{fox1961H-function} as the generalisation of the Meijer $G$-function.

Here, we provide the definition and properties of the Fox $H$-function that are needed to describe the gNFW profile as a gravitational lens. For a more comprehensive description of the Fox H-function, refer to e.g., \cite{kilbas1999, kilbas2004h}. The Fox H-function is in general a complex-valued function defined by the Mellin-Barnes integral
\begin{align}\label{eq: Fox H-function}
\displaystyle
\nonumber
H_{p,q}^{m,n}(x)&=H_{p,q}^{m,n} \!\left[ x \left| \begin{matrix}
( a_1 , A_1 ),\ldots,( a_p , A_p ) \\
( b_1 , B_1 ),\ldots,( b_q , B_q ) \end{matrix} \right. \right]\\
&= \dfrac{1}{2\pi i}\int_{\mathcal{L}} \Theta(u)x^{-u}du
\end{align}
with 
\begin{equation}\label{eq: Fox H-function argument}
 \Theta(u)=\dfrac{\displaystyle \prod_{j=1}^{m}\Gamma(b_j+B_ju) \prod_{j=1}^{n}\Gamma(1-a_j-A_j u)}{\displaystyle  \prod_{j=n+1}^{p}\Gamma(a_j+A_j u)\prod_{j=m+1}^{q}\Gamma(1-b_j -B_j u)}, 
\end{equation}
where $m,n,p,q \in\mathbb{N}_{0}$, $A_j\,(j=1,\ldots,p)$ and $B_j\,(j=1,\ldots,q)$ are positive real numbers, while $a_j\,(j=1,\ldots,p)$ and $b_j\,(j=1,\ldots,q)$ are complex numbers. From the properties of the $\Gamma$ function, we have that the poles present in \eqref{eq: Fox H-function argument}  correspond to the points where the $\Gamma$ functions in its numerator diverge, which occurs when their argument equals zero or a negative integer. Therefore, the contour $\mathcal{L}$ separates the poles of the product $\prod_{j=1}^{m}\Gamma(b_j+B_ju)$ from those of the product $\prod_{j=1}^{n}\Gamma(1-a_j-A_j u)$. In the special case $A_j=1$  and $B_j=1$, the Fox $H$-function reduces to the Meijer $G$-function
\begin{align}\label{eq: Fox H-function}
\displaystyle
\nonumber
G_{p,q}^{m,n}(x) = G_{p,q}^{m,n} \!\left[ x \left| \begin{matrix}
a_1 ,\ldots, a_p\\
b_1 ,\ldots, b_q\end{matrix} \right. \right].
\end{align}

Following \cite{kilbas1999}, we evaluate the Fox $H$-function using its power series expansion. This series expansion depends on the multiplicity of the different poles $\beta_j$ (of functions $\Gamma(b_j+B_j u))$ and poles $\alpha_j$ (of functions $\Gamma(1-a_j-A_j u)$), which are given as
\begin{equation}\label{eq: poles beta}
\beta_{j}= -\left(\dfrac{b_j+k_j}{B_j}\right)\quad (j=1,\dots,m\quad \text{and}\quad k_j\in\mathbb{N}_{0})
\end{equation}
and
\begin{equation}\label{eq: poles alpha}
\alpha_{j}=\dfrac{1-a_j+l_j}{A_j}\quad(j=1,\dots,n\quad \text{and}\quad l_j\in\mathbb{N}_{0}).
\end{equation}

Considering the parameters 
\begin{equation}\label{eq: convergence conditions H}
\Delta = \sum_{j=1}^{q}B_j -  \sum_{j=1}^{p}A_j \quad \text{and}\quad \delta = \prod_{j=1}^{q}B_j^{B_j}\prod_{j=1}^{p}A_j^{-A_j},
\end{equation}
the power series expansion converges when (see Theorems 2 - 6 in \cite{kilbas1999})
\begin{subequations}\label{cases}
\begin{align}
&\text{\textbf{Case 1:} $\Delta>0$ and  $x\neq 0$, or $\Delta=0$ and  $0<|x|<\delta$.}\label{case 1}\\
&\text{\textbf{Case 2:} $\Delta<0$ and  $x\neq 0$, or $\Delta=0$ and  $|x|>\delta$.\label{case 2}}
\end{align}
\end{subequations}
Thus, in both cases, the power series expansion of the Fox $H$-function can be expressed as
\begin{align}\label{eq: H series general}
\nonumber
&H_{p,q}^{m,n}(x)=\underbrace{\sum_i\sum_{k_i}  \mathcal{C}_{1i} x^{-u_{pi}}}_\text{Simple poles}\\\nonumber
&+\underbrace{\sum_{k_1}\mathcal{C}_{2}x^{-u_{p1}}\left(\mathcal{C}^{\prime}_{2}-\ln(x)\right)}_{\text{Second order poles}}\\
&+\underbrace{\sum_{k_1}\mathcal{C}_{3}x^{-u_{p1}}\left(\dfrac{1}{2}\mathcal{C}^{\prime 2}_{3}+\dfrac{1}{2}\mathcal{C}^{\prime\prime}_{3}-\mathcal{C}^{\prime}_{3}\ln(x) + \dfrac{1}{2}\ln^2(x)\right)}_{\text{Third order poles}},
\end{align}
where we have considered the expansion for poles up to third order. In particular, we make explicit the possibility of having several $\Gamma$ functions with simple poles, while for higher order poles we only consider the possibility of having one pair or triplet of $\Gamma$ functions with poles of second or third order, respectively. This is what we face in this work, however, for poles of second or third order, if more than one pair or triplet of $\Gamma$ functions with poles of these orders appear, we can add a new summation (one for each extra pair or triplet) with the same functional form as in \eqref{eq: H series general}. \cite{kilbas1999} provide the general expansion for poles of arbitrary order, which reduces to \eqref{eq: H series general} after some algebra.

The form taken by the coefficients $\mathcal{C}_{s}$, $\mathcal{C}_{s}^{\prime}$, and $\mathcal{C}_{s}^{\prime\prime}$ (with $s=1, 2,3$ as appropriate) in \eqref{eq: H series general}, depends on each of the aforementioned cases, which we make explicit below. Here, the prime does not denote derivative.

\textbf{Case 1 \eqref{case 1}:} In this case the poles of interest are those given by \eqref{eq: poles beta}. At most $m$ of them are simple poles, which we denote $u_{pi}=\beta_i$ ($i=1,\dots,m$ at most). The corresponding coefficients take the form
\begin{equation}
\mathcal{C}_{1i} = \dfrac{(-1)^{k_i}}{B_i k_i!}\dfrac{\displaystyle \prod_{\substack{j=1 \\ j\neq i}}^{m}\Gamma(b_j+B_j\beta_i) \prod_{j=1}^{n}\Gamma(1-a_j-A_j \beta_i)}{\displaystyle  \prod_{j=n+1}^{p}\Gamma(a_j+A_j \beta_i)\prod_{j=m+1}^{q}\Gamma(1-b_j -B_j \beta_i)}.
\end{equation}

For poles of  higher order we have $u_{p1}=\beta_1$, where $\beta_1=\beta_2$ for poles of second order and $\beta_1=\beta_2=\beta_3$ for poles of third order. These relations allow us to write $k_2$ and $k_3$ in terms of $k_1$, such that, for $s=2,3$ the corresponding coefficients are
\begin{equation}
\mathcal{C}_{s} = \left(\prod_{j=1}^{s}\dfrac{(-1)^{k_j}}{B_{j}k_j!}\right)\dfrac{\displaystyle \prod_{j=s+1}^{m}\Gamma(b_j+B_j \beta_1) \prod_{j=1}^{n}\Gamma(1-a_j-A_j \beta_1)}{\displaystyle  \prod_{j=n+1}^{p}\Gamma(a_j+A_j\beta_1)\prod_{j=m+1}^{q}\Gamma(1-b_j -B_j \beta_1)},
\end{equation}
\begin{align}
\nonumber
\mathcal{C}^{\prime}_{s} &= \sum_{j=s+1}^{m}B_j\psi_{0}(b_j+B_j \beta_1) - \sum_{j=1}^{n}A_j\psi_{0}(1-a_j-A_j\beta_1)\\\nonumber
& - \sum_{j=n+1}^{p}A_j\psi_{0}(a_j+A_j \beta_1) + \sum_{j=m+1}^{q}B_j\psi_{0}(1-b_j-B_j\beta_1)\\
& +  \sum_{j=1}^{s}B_j\psi_{0}(1+k_j),
\end{align}
and
\begin{align}
\nonumber
\mathcal{C}^{\prime\prime}_{3} &= \sum_{j=4}^{m}B_j^2\psi_{1}(b_j+B_j \beta_1) + \sum_{j=1}^{n}A_j^2\psi_{1}(1-a_j-A_j\beta_1)\\\nonumber
& - \sum_{j=n+1}^{p}A_j^2\psi_{1}(a_j+A_j \beta_1) - \sum_{j=m+1}^{q}B_j^2\psi_{1}(1-b_j-B_j \beta_1)\\
& +  \sum_{j=1}^{3}B_j^2\left(\dfrac{\pi^2}{3}-\psi_{1}(1+k_j)\right).
\end{align}

\textbf{Case 2 \eqref{case 2}:} In this case the poles of interest are those given by \eqref{eq: poles alpha}. At most $n$ of them are simple poles, which we denote $u_{pi}=\alpha_i$ ($i=1\dots n$ at most). The corresponding coefficients take the form
\begin{equation}
\mathcal{C}_{1i} = \dfrac{(-1)^{l_i}}{A_i l_i!}\dfrac{\displaystyle \prod_{j=1}^{m}\Gamma(b_j+B_j\alpha_i) \prod_{\substack{j=1 \\ j\neq i}}^{n}\Gamma(1-a_j-A_j \alpha_i)}{\displaystyle  \prod_{j=n+1}^{p}\Gamma(a_j+A_j \alpha_i)\prod_{j=m+1}^{q}\Gamma(1-b_j -B_j \alpha_i)},
\end{equation}

For poles of higher order we have $u_{p1}=\alpha_1$, where $\alpha_1=\alpha_2$ for poles of second order and $\alpha_1=\alpha_2=\alpha_3$ for poles of third order. These relations allow us to write $l_2$ and $l_3$ in terms of $l_1$, such that, for $s=2,3$ the corresponding coefficients are 
\begin{equation}
\mathcal{C}_{s} = -\left(\prod_{j=1}^{s}\dfrac{(-1)^{l_j-1}}{A_{j}l_j!}\right)\dfrac{\displaystyle \prod_{j=1}^{m}\Gamma(b_j+B_j \alpha_1) \prod_{j=s+1}^{n}\Gamma(1-a_j-A_j \alpha_1)}{\displaystyle  \prod_{j=n+1}^{p}\Gamma(a_j+A_j\alpha_1)\prod_{j=m+1}^{q}\Gamma(1-b_j -B_j \alpha_1)},
\end{equation}
\begin{align}
\nonumber
\mathcal{C}^{\prime}_{s} &= \sum_{j=1}^{m}B_j\psi_{0}(b_j+B_j \alpha_1) - \sum_{j=s+1}^{n}A_j\psi_{0}(1-a_j-A_j\alpha_1)\\\nonumber
& - \sum_{j=n+1}^{p}A_j\psi_{0}(a_j+A_j \alpha_1) + \sum_{j=m+1}^{q}B_j\psi_{0}(1-b_j-B_j\alpha_1)\\
& -  \sum_{j=1}^{s}A_j\psi_{0}(1+l_j),
\end{align}
and
\begin{align}
\nonumber
\mathcal{C}^{\prime\prime}_{3} &= \sum_{j=1}^{m}B_j^2\psi_{1}(b_j+B_j \alpha_1) + \sum_{j=4}^{n}A_j^2\psi_{1}(1-a_j-A_j\alpha_1)\\\nonumber
& - \sum_{j=n+1}^{p}A_j^2\psi_{1}(a_j+A_j \alpha_1) - \sum_{j=m+1}^{q}B_j^2\psi_{1}(1-b_j-B_j \alpha_1)\\
& +  \sum_{j=1}^{3}A_j^2\left(\dfrac{\pi^2}{3}-\psi_{1}(1+l_j)\right).
\end{align}

As a final remark, when working with the power series \eqref{eq: H series general}, it s useful to keep in mind that
\begin{equation}
\lim_{x\to -k}\dfrac{1}{\Gamma(1+x)}=\dfrac{1}{\Gamma(1-k)}=0,
\end{equation}
\begin{equation}
\lim_{x\to -k}\dfrac{\psi_
0(1+x)}{\Gamma(1+x)} = \dfrac{\psi_
0(1-k)}{\Gamma(1-k)} = (-1)^{k}\Gamma(k),
\end{equation}
\begin{equation}
\lim_{x\to -k}\dfrac{\psi_1(1+x)}{\Gamma(1+x)^2} = \dfrac{\psi_1(1-k)}{\Gamma(1-k)^2} = \Gamma(k)^2,
\end{equation}
with $k\in \mathbb{Z}^{+}$.

\section{A simple application case}\label{appendix: sis and nis}
Consider a Singular Isothermal Sphere (SIS) whose mass density is given by
\begin{equation}\label{eq: sis mass distribution}
\rho_{\text{SIS}}(r) = \dfrac{\sigma_v^2}{2\pi G r^2}, 
\end{equation}
where $\sigma_v$ is the velocity dispersion, and $G$ is the gravitational constant. Its corresponding surface mass density, given in terms of the Abel integral \eqref{eq: surface mass density abel} (in analogy to \eqref{eq: gnfw surface mass density abel}) reads 
\begin{equation}\label{eq: sis surface mass density}
\Sigma_{\text{SIS}}(x)=\dfrac{\sigma^2_v}{\pi G \xi_0}\int_{x}^{\infty}\dfrac{dt}{t\sqrt{t^2-x^2}},
\end{equation}
with $t=r/\xi_0$, where in this case the normalisation scale length $\xi_0$ could be, for example, the Einstein radius. When we write \eqref{eq: sis surface mass density} as the Mellin convolution \eqref{eq: mellin convolution} between $f_{1, \text{SIS}}$ and $f_{2, \text{SIS}}$, we have that
\begin{equation}\label{eq: sis f1}
f_{1, \text{SIS}}(h=t)=\dfrac{\sigma^2_v}{\pi G \xi_0},
\end{equation}
for which its Mellin transform $\mathfrak{M}_{f_1, \text{SIS}}$ (see \eqref{eq: mellin transform}) does not converge. This result is enough to conclude that \eqref{eq: sis surface mass density} cannot be written as \eqref{eq: mellin transform technique}, and, in consequence, it cannot be given in terms of the Fox $H$-Function. At least, not by this mean.

We can add a core with radius $r_c$ by means of $r^2\to r^2+r^2_c$, forming a Non-Singular Isothermal Sphere (NIS) whose mass density is
\begin{equation}\label{eq: nis distribution}
\rho_{\text{NIS}}(r) = \dfrac{\sigma_v^2}{2\pi G \left(r^2+r_{c}^2\right)}.    
\end{equation}
This mass distribution is simple enough to be described as a gravitational lens by direct integration, as shown in \cite{hinshaw_nis1987}. For this reason, it is a good exercise to describe it in terms of the Fox $H$-function. In particular, doing this exercise helps us  to become familiar with the series expansion \eqref{eq: H series general}.

Following the same process that we applied for the gNFW in Sec. \ref{sec: gNFW lensing}, and by taking the normalisation scale length $\xi_0=r_c$, it can be shown without much trouble that for the NIS, its surface mass density can be written as the Mellin convolution (see \eqref{eq: mellin convolution}) between
\begin{equation}
f_{1,\text{NIS}}(t=b)=\dfrac{\sigma^2_v}{\pi G r_c}\dfrac{t^2}{t^2+1}
\end{equation}
and $f_{2,\text{NIS}}$, which turns out to be identical to \eqref{eq: gNFW f2}. Their corresponding Mellin transforms (see \eqref{eq: mellin transform}) are 
\begin{equation}
\mathfrak{M}_{f_1,\,\text{NIS}}(u)= -\dfrac{\sigma^2_v}{2\pi G r_c}\Gamma\left(1-\frac{u}{2}\right)\left(\frac{u}{2}\right)
\end{equation}
and \eqref{eq: gNFW mellin transform 2}, respectively. Therore, from \eqref{eq: mellin transform technique} and using the change of variable $v=1+u/2$ (for convenience), the surface mass density becomes
\begin{equation}\label{eq: nis surface mass density Mellin}
\Sigma_{\text{NIS}}(x) = \Sigma_0^{\text{NIS}} x
\left(\dfrac{1}{2\pi i}
\int_{\mathcal{L}}\Theta_{\Sigma}^{\text{NIS}}(v)x^{-2v}dv\right),
\end{equation}
with
\begin{equation}\label{eq: nis theta surface mass density}
\Theta_{\Sigma}^{\text{NIS}}(v)= \Gamma\left(-\frac{1}{2}+v\right)\Gamma(1-v)
\end{equation}
and
\begin{equation}
\Sigma_0^{\text{NIS}}=\dfrac{\sigma_{v}^2}{2\sqrt{\pi} G r_c},
\end{equation}
where it is clear that \eqref{eq: nis surface mass density Mellin} satisfies \eqref{eq: general surface mass density} for $a=1$ and $b=2$. Thus, from \eqref{eq: nis theta surface mass density} along with the outline given in Sec. \ref{sec: general result}, we can write the surface mass density, projected mass, and deflection potential explicitly in terms of the Fox $H$-function \eqref{eq: Fox H-function} as
\begin{equation}\label{eq: nis surface mass density H}
\displaystyle
\Sigma_{\text{NIS}}(x)=\Sigma_0^{\text{NIS}} x H_{1,1}^{1,1} \!\left[ x^2 \left| \begin{matrix}
( 0 , 1 ) \\
( -\frac{1}{2} , 1 ) \end{matrix} \right. \right],
\end{equation}
\begin{equation}\label{eq: nis projected mass H}
M_{\text{NIS}}(x) = \pi r_c^2\Sigma_0^{\text{NIS}}x^{3}
H^{1,2}_{2,2} \!\left[ x^2 \left| \begin{matrix}
( 0 , 1 ),(-\frac{1}{2},1) \\
( -\frac{1}{2} , 1 ),(-\frac{3}{2},1) \end{matrix} \right. \right],
\end{equation}
\begin{equation}\label{eq: nis deflection potential H}
\psi_{\text{NIS}}(x) = \dfrac{\kappa_0^{\text{NIS}} x^{3}}{2}
H^{1,3}_{3,3} \!\left[ x^2 \left| \begin{matrix}
( 0 , 1 ),(-\frac{1}{2},1),(-\frac{1}{2},1) \\
( -\frac{1}{2} , 1 ),(-\frac{3}{2},1),(-\frac{3}{2},1) \end{matrix} \right. \right],
\end{equation}
respectively.

From parameters \eqref{eq: convergence conditions H}, we can show that the three Fox $H$-functions in \eqref{eq: nis surface mass density H}-\eqref{eq: nis deflection potential H} satisfy $\Delta = 0$ and $\delta = 1$. Hence, as described in \ref{cases}, they converge for $0<x^2<1$ and $x^2>1$. Therefore, considering that for us $x>0$ , and by applying the power series representation \eqref{eq: H series general}, we get 
\begin{equation}\label{eq: nis surface mass density series}
\Sigma_{\text{NIS}}(x) = \Sigma_0^{\text{NIS}}
\begin{dcases}
\displaystyle\sum_{k=0}^{\infty}\dfrac{(-1)^{k}}{k!}\Gamma\left(\frac{1}{2}+k\right)x^{2k} \hspace*{\fill}\qquad\text{if }0<x<1\\
\displaystyle\sum_{l=0}^{\infty}\dfrac{(-1)^{l}}{l!}\Gamma\left(\frac{1}{2}+l\right)x^{-2l-1} \hspace*{\fill}\qquad\text{if } x> 1
\end{dcases}.
\end{equation}
\begin{align}\label{eq: nis projected mass series}
\nonumber
&M_{\text{NIS}}(x) = \pi r_c^2\Sigma_0^{\text{NIS}}\\
&\times
\begin{dcases}
\displaystyle\sum_{k=0}^{\infty}\dfrac{(-1)^{k}}{(1+k)!}\Gamma\left(\frac{1}{2}+k\right)x^{2k+2} \hspace*{\fill}\qquad\text{if } 0<x<1\\
\displaystyle 2\sum_{l=0}^{\infty}\dfrac{(-1)^{l}}{l!}\Gamma\left(\frac{1}{2}+l\right)\frac{x^{-2l+1}}{(1-2l)} - 2\sqrt{\pi} \hspace*{\fill}\qquad\text{if } x> 1
\end{dcases},
\end{align}
\begin{align}\label{eq: nis deflection potential series}
\nonumber
&\psi_{\text{NIS}}(x) = \frac{\kappa_0^{\text{NIS}}}{2}\\
&\times
\begin{dcases}
\displaystyle\sum_{k=0}^{\infty}\dfrac{(-1)^{k}}{k!}\Gamma\left(\frac{1}{2}+k\right)\frac{x^{2k+2}}{(1+k)^2} \hspace*{\fill}\qquad\text{if } 0<x<1\\[3.5ex]
\displaystyle 4\sum_{l=0}^{\infty}\dfrac{(-1)^{l}}{l!}\Gamma\left(\frac{1}{2}+l\right)\frac{x^{-2l+1}}{(1-2l)^2} - 4\sqrt{\pi}\left(\ln\left(\dfrac{x}{2}\right)+1\right) & \\
\hspace*{\fill}\qquad\text{if } x> 1
\end{dcases}.
\end{align}
The discontinuities at $x=0$ and $x=1$ in \eqref{eq: nis surface mass density series}-\eqref{eq: nis deflection potential series} are removable, so that for $x\geq 0$ it can be shown that they converge to
\begin{equation}
\Sigma_{\text{NIS}}(x) = \dfrac{\sqrt{\pi}\Sigma_0^{\text{NIS}}}{\sqrt{1+x^2}},
\end{equation}
\begin{equation}
M_{\text{NIS}}(x) = 2\pi^{3/2} r_c^2 \Sigma_0^{\text{NIS}} \left(\sqrt{1+x^2} -1 \right),
\end{equation}
\begin{equation}
\psi_{\text{NIS}}(x) = 2\sqrt{\pi} \kappa_{0}^{\text{NIS}}\left(\sqrt{1+x^2} - \ln\left(\frac{1+\sqrt{1+x^2}}{2}\right) -1 \right),
\end{equation}
respectively. For this application case, it is worth noting that in all cases the Fox $H$-function coincides with the Meijer $G$-function. From these results, and using definitions \eqref{eq: shear} and \eqref{eq: mean shear}, we can write the corresponding shear and mean shear as
\begin{equation}
\gamma_{\text{NIS}}=\sqrt{\pi}\kappa_0\left(\dfrac{2+x^2-2\sqrt{1+x^2}}{x^2\sqrt{1+x^2}}\right)
\end{equation}
and
\begin{equation}
\overline{\gamma}_{\text{NIS}}(x) = \dfrac{2\sqrt{\pi} \kappa_{0}^{\text{NIS}}}{x^2}\left(\sqrt{1+x^2} - 2\ln\left(\frac{1+\sqrt{1+x^2}}{2}\right) -1 \right),
\end{equation}
respectively. Similarly, from \eqref{eq: deflection angle} the deflection angle takes the form
\begin{equation}
\alpha_{\text{NIS}}(x)=\dfrac{2\sqrt{\pi}\kappa_{0}^{\text{NIS}}}{x}\left(\sqrt{1+x^2}-1\right), 
\end{equation}
where $\alpha_{\text{NIS}}(0)=0$.

\section{NFW and Hernquist}\label{appendix: nfw and hernquist}
In this appendix, for the NFW and Hernquist profiles we list their analytical properties as gravitational lenses (see e.g, \cite{bartelmann1996NFWarcs, keeton2001catalog}). For this task, we use the auxiliary functions
\begin{equation}
f(x) =
\begin{dcases}
\arccosh{\left(\dfrac{1}{x}\right)}\hspace*{\fill}\qquad\text{if } 0 < x < 1\\
\arccos{\left(\dfrac{1}{x}\right)}\hspace*{\fill}\qquad\text{if } x> 1
\end{dcases}
\end{equation}
and
\begin{equation}
s(x) =
\begin{dcases}
1\hspace*{\fill}\qquad\text{if } 0 < x < 1\\
-1\hspace*{\fill}\qquad\text{if } x> 1
\end{dcases},
\end{equation}
which help us to get compact expressions. It can be shown that with the appropriate value of $n$, \eqref{eq: gNFW surface mass density H}-\eqref{eq: gNFW deflection potential H} along with \eqref{eq: deflection angle}, \eqref{eq: shear} and \eqref{eq: mean shear - potential} reduce to the expressions given below. Note that, with the exception of the deflection angle, these expressions are valid for $x>0$.   

\subsection{NFW}
Here, the characteristic surface mass density is $\Sigma_0^{\text{NFW}} = 2\sqrt{\pi}\rho_0 r_0 $, where $\rho_0$ is a characteristic density and $r_0$ is a scale length. With this, we can write
\begin{equation}\label{eq: nfw surface mass density}
\Sigma_{\text{NFW}}(x) = \dfrac{\Sigma_0^{\text{NFW}}}{\sqrt{\pi}}
\begin{dcases}
\dfrac{1}{3}& \text{if } x= 1\\
\dfrac{1}{x^2-1}\left(1 - \frac{f(x)}{\sqrt{|1-x^2|}}\right) & \text{if } x\neq 1
\end{dcases},
\end{equation}
\begin{equation}\label{eq: nfw projected mass}
M_{\text{NFW}}(x) = 2\sqrt{\pi}r_0^2\Sigma_0^{\text{NFW}}
\begin{dcases}
1-\ln(2)& \text{if } x= 1\\
\ln\left(\dfrac{x}{2}\right) + \frac{f(x)}{\sqrt{|1-x^2|}}& \text{if } x\neq 1
\end{dcases},
\end{equation}
\begin{equation}\label{eq: nfw deflection potential}
\psi_{\text{NFW}}(x) = \dfrac{\kappa_0^{\text{NFW}}}{\sqrt{\pi}}
\begin{dcases}
\ln^2\left(2\right) & \text{if } x= 1\\
\ln^2\left(\dfrac{x}{2}\right)  -s(x)f^2(x) & \text{if } x\neq 1
\end{dcases},
\end{equation}
\begin{align}\label{eq: nfw shear}
\nonumber
&\gamma_{\text{NFW}}(x) = \kappa_0^{\text{NFW}}/(\sqrt{\pi}x^2)\\
&\times
\begin{dcases}
\dfrac{5}{3}-2\ln\left(2\right) & \text{if } x= 1\\
2\ln\left(\dfrac{x}{2}\right) + \dfrac{x^2}{1-x^2}+\dfrac{(2-3x^2)s(x)f(x)}{(|1-x^2|)^{3/2}} & \text{if } x\neq 1
\end{dcases},
\end{align}
\begin{align}\label{eq: nfw mean shear}
\nonumber
&\overline{\gamma}_{\text{NFW}}(x) = 2\kappa_0^{\text{NFW}}/(\sqrt{\pi}x^2)\\
&\times
\begin{dcases}
\ln^2\left(2\right)+\ln\left(2\right)-1 & \text{if } x= 1\\
\ln^2\left(\dfrac{x}{2}\right)-\ln\left(\dfrac{x}{2}\right) -s(x)f^2(x)-\dfrac{f(x)}{\sqrt{|1-x^2|}}  & \text{if } x\neq 1
\end{dcases},
\end{align}
and
\begin{equation}\label{eq: nfw deflection angle}
\alpha_{\text{NFW}}(x) = \dfrac{2\kappa_0^{\text{NFW}}}{\sqrt{\pi} x}
\begin{dcases}
1-\ln(2)& \text{if } |x|= 1\\
\ln\left(\dfrac{|x|}{2}\right) + \frac{f(|x|)}{\sqrt{|1-x^2|}}& \text{if } |x|\neq 1
\end{dcases},
\end{equation}
where $\alpha_{\text{NFW}}(0)=0$.

\subsection{Hernquist}
Here, the characteristic surface mass density is $\Sigma_0^{\text{H}} = M_{\infty}/(2\sqrt{\pi}r_0^2)$, where $M_{\infty}$ is the total mass of the distribution and $r_0$ is a scale length. With this, we can write
\begin{equation}\label{eq: h surface mass density}
\Sigma_{\text{H}}(x) = \dfrac{\Sigma_0^{\text{H}}}{\sqrt{\pi}}
\begin{dcases}
\dfrac{4}{15}& \text{if } x= 1\\
\dfrac{1}{(x^2-1)^2}\left(\frac{(2+x^2)f(x)}{\sqrt{|1-x^2|}}-3\right) & \text{if } x\neq 1
\end{dcases},
\end{equation}
\begin{equation}\label{eq: h projected mass}
M_{\text{H}}(x) = 2\sqrt{\pi}r_0^2\Sigma_0^{\text{H}}
\begin{dcases}
\dfrac{1}{3}& \text{if } x= 1\\
\dfrac{x^2}{x^2-1}\left(1 - \frac{f(x)}{\sqrt{|1-x^2|}}\right) & \text{if } x\neq 1
\end{dcases},
\end{equation}
\begin{equation}\label{eq: h deflection potential}
\psi_{\text{H}}(x) = \dfrac{2\kappa_0^{\text{H}}}{\sqrt{\pi}}
\begin{dcases}
1-\ln\left(2\right) & \text{if } x= 1\\
\ln\left(\dfrac{x}{2}\right) + \dfrac{f(x)}{\sqrt{|1-x^2|}}  & \text{if } x\neq 1
\end{dcases},
\end{equation}
\begin{align}\label{eq: h shear}
\gamma_{\text{H}}(x) = \dfrac{\kappa_0^{\text{H}}}{\sqrt{\pi}}
\begin{dcases}
\dfrac{2}{5}& \text{if } x= 1\\
\dfrac{1+2x^2}{(x^2-1)^2}-\dfrac{3x^2f(x)}{(|1-x^2|)^{5/2}}& \text{if } x\neq 1
\end{dcases},
\end{align}
\begin{align}\label{eq: h mean shear}
\nonumber
&\overline{\gamma}_{\text{H}}(x) = 2\kappa_0^{\text{H}}/(\sqrt{\pi}x^2)\\
&\times
\begin{dcases}
\dfrac{5}{3}-2\ln\left(2\right) & \text{if } x= 1\\
2\ln\left(\dfrac{x}{2}\right) +\dfrac{x^2}{1-x^2}+\dfrac{(2-3x^2)s(x)f(x)}{(|1-x^2|)^{3/2}} & \text{if } x\neq 1
\end{dcases},
\end{align}
and
\begin{equation}\label{eq: h deflection angle}
\alpha_{\text{H}}(x) = \dfrac{2\kappa_0^{\text{H}}}{\sqrt{\pi} x}
\begin{dcases}
\dfrac{1}{3}& \text{if } |x|= 1\\
\dfrac{x^2}{x^2-1}\left(1 - \frac{f(|x|)}{\sqrt{|1-x^2|}}\right) & \text{if } |x|\neq 1
\end{dcases},
\end{equation}
where $\alpha_{\text{H}}(0)=0$.

\bsp	
\label{lastpage}
\end{document}